\documentclass[preprint,12pt]{elsarticle}

\usepackage{graphicx}
\graphicspath{{pics/}}
\usepackage{amssymb}

\journal{ }

\begin{document}

\begin{frontmatter}

\title{Is there a Relationship between the Elongational Viscosity and the First Normal Stress Difference in Polymer Solutions?}

\author{A.~Zell $^1$, S.~Gier $^1$, S.~Rafa\"{\i} $^2$, and C.~Wagner $^{1,*}$}
\address{$^1$ Technische Physik, Universit\"at des Saarlandes, Postfach 151150, 66041
Saarbr\"ucken, Germany\\ $^2$ Complex Fluid Dynamics Group, Laboratoire de Spectrometrie Physique (CNRS-UJF), 38402 St Martin d'Heres, France\\
$^*$ c.wagner at mx.uni-saarland.de}

\begin{abstract}
We investigate a variety of different polymer solutions in shear and
elongational flow. The shear flow is created in the
cone-plate-geometry of a commercial rheometer. We use capillary
thinning of a filament that is formed by a polymer solution in the
Capillary Breakup Extensional Rheometer (CaBER) as an elongational
flow. We compare the relaxation time and the elongational viscosity
measured in the CaBER with the first normal stress difference and
the relaxation time that we measured in our rheometer. All of these
four quantities depend on different fluid parameters - the viscosity
of the polymer solution, the polymer concentration within the
solution, and the molecular weight of the polymers - and on the
shear rate (in the shear flow measurements). Nevertheless, we find
that the first normal stress coefficient depends quadratically on
the CaBER relaxation time. A simple model is presented that explains
this relation.
\end{abstract}


%\vskip 0.2 truein

%\pacs{??pacs??}

\end{frontmatter}

%\maketitle

%\vskip2pc

%-------------------- SEC I -----------------------------
\section{Introduction}
\label{sec1}
%--------------------------------------------------------

Viscoelasticity is probably the most prominent Non-Newtonian effect;
it manifests itself a high elongational viscosity and a non zero
first normal stress difference. Some examples of these effects
are the filaments one can make with saliva or dye swell, i.e., a
viscoelastic solution emerges from a pipe in the form of a jet with a
width larger than the pipe's width. Both of these effects are
believed to be due to the same microscopic reason, the resistance to
stretching of semi-flexible polymers. However, to our knowledge,
there have been no experimental attempts to relate these two accessible
quantities in a systematic manner.

The elongational viscosity of polymer solutions can be several
orders of magnitude larger than the solvent viscosity, especially
for flexible high molecular polymers. In contrast, the
shear viscosity remains on the order of the solvent viscosity. Shear
flow can be divided into a rotational part and an elongational part. The
elongational part stretches the macromolecules and induces stress
while the rotational part leads to a tumbling of the polymers, and
the direction of stress gets averaged out. In pure elongational
flow, the orientation of the stretching remains constant and the polymers get
stretched more and more until, eventually, they are elongated to their
contour length, and the elongational viscosity reaches a plateau. In
many applications, elongational flow affects the processability more
severely, e.g., in fiber spinning, spraying, deposition of
pesticides, etc. where the flow is elongational, at least to a large
extent.

However, the definition of the first Normal stress $N_1 =
\tau_{xx}-\tau_{yy}$ (with $\tau_{ab}$ the components of the stress
tensor) in shear flow compared to the definition of the elongational
viscosity
$\eta_e=\left(\tau_{xx}-\tau_{yy}\right)/\dot{\varepsilon}$ in
elongational flow suggests the existence of a direct relation between
these two quantities ($\dot{\varepsilon}$ is the elongational rate).
>From an experimental point of view, measurement of the
elongational viscosity of dilute polymer solutions is a nontrivial
task and only recently a method called CaBER (Capillary Breakup
Extensional Rheometer) became available. A droplet is placed between
two plates, and after they are separated by a linear motor the
capillary bridge between them starts to shrink. Instead of break up, a
cylindrical filament is formed due to the high viscoelastic
stresses. By balancing surface tension and viscoelastic stresses an
apparent elongational viscosity can be deduced by simply measuring
the thickness versus time $h(t)$ of the shrinking filament. In
such a set-up, the elongational rate is chosen by the system and can
not be controlled.

The relation between shear and elongational flows has been previously investigated.
A principle approach was to determine the characteristic
time constant of the filament in a CaBER setup that shrinks
exponentially in time, and to compare it with the respective shear
quantities (shear viscosity and relaxation time) using different
models (typically, Zimm~\cite{Zimm56} or Rouse~\cite{Rouse53}).
Gupta et al.~\cite{Gupta00} performed shear and oscillatory shear
experiments, which they compared to extensional rheological data with
a filament-stretching device~\cite{Sridhar91,Tirtaatmadja93}. The
samples were different polystyrene solutions. They found that the
shear properties could be well described by the Zimm model, whereas
the extensional data show a direct dependence of the stress growth
on concentration and molecular weight of the polymers; this agrees
with the Rouse model. Nevertheless, by using a FENE-P model with
Zimm parameters they were not able to predict the results or to fit
the extensional data.

In 2003, Lindner et al.~\cite{Lindner03} tried to compare the
elastic properties in elongational and shear flow by
using an opposed nozzle rheometer \cite{Fuller87} and a classical
rotational rheometer, respectively. They explicitly tried to fit
their rheometric measurements with the FENE-P model, but, to some
extent, they were only able to fit the normal stress data. The
shear thinning effects in their set of highly elastic solutions were
not covered by the FENE parameters obtained from the normal stress
data. Still, they could calculate the elongational viscosity from
the FENE parameters and compare them to the measured elongational
viscosity by adjusting the finite extensibility parameter $b$.

Plog et al.~\cite{Plog05} used the technique of capillary thinning
rheometry to characterize the dependence of the CaBER relaxation
time on the molar mass distribution of the polymers in solution.
They found good agreement between the molecular weight
distribution obtained from the CaBER measurements and experiments on
size-exclusion-chromatography, multi angle laser light scattering
and differential refractometry, but they were not able to correlate
the results with standard rheometric measurements.

In 2006, Clasen et al.~\cite{Clasen06} published their work
concerning the question of the determination of the overlap
concentration of polymer solutions in CaBER experiments. They
compared CaBER and small amplitude oscillatory shear (SAOS)
measurements on polystyrene solutions below the "classical" overlap
concentration. For the shear measurements, they found that the
relaxation times of their solutions agree very well with the Zimm
relaxation times and show only a slight increase when approaching
the overlap concentration. The CaBER measurements, however, revealed
quite different behavior: at low concentrations the relaxation
times are below the Zimm values, but they continuously increase with
increasing concentration, so that, for concentrations near the
overlap concentration, the values are much higher. Therefore, they
concluded that there must be a so called critical polymer
concentration in elongational measurements, which is orders of
magnitudes smaller than the overlap concentration that is found in
shear experiments. The result of an increase in the relaxation time
with increasing concentration, even below the overlap concentration,
in an elongation experiment was also observed by Tirtaatmadja et
al.~\cite{Tirtaatmadja06} and by Amarouchene et
al.~\cite{Amarouchene01}, who both investigated the droplet
detachment of polyethyleneoxide in glycerol/water mixtures or pure
water, respectively.

Here, we compare the first normal stress coefficient, which we
determined from rheometer measurements, with the relaxation time,
which we measure in CaBER experiments. Thereby, we can directly
relate the first normal stress difference to the elongational
viscosity in polymer solutions. We find that the first normal stress coefficient depends quadratically on the CaBER relaxation time.

In section \ref{sec2}, we discuss the theoretical background on the
basis of different polymer models. In section \ref{sec3} the
experimental methods are presented. In section \ref{sec5} we report
our measurements and results. Conclusions are drawn in section
\ref{sec6}.

%-------------------- SEC II -----------------------------
\section{Theoretical Background}
\label{sec2}
%--------------------------------------------------------

Despite the numerous theoretical contributions to the field, a full
quantitative modeling of viscoelastic properties of complex fluids
remains a challenge nowadays. However, standard models have been
proven to describe important features of polymeric solutions within
some limits. Among them, Oldroyd-B is a minimal model giving rise
to viscoelastic effects. It consists of considering two beads
connected by a Hookean spring suspended in an incompressible
Newtonian fluid. The linear spring force puts no limit on the extent
to which the dumbbell can be stretched. The FENE-P model (finitely
extensible non-linear elastic) corrects for this unphysical
behavior. We compared our data to those two models.

\subsection{Oldroyd-B model}

Based on the microscopic picture of an elastic dumbbell, the
Oldroyd-B model is probably the simplest linear viscoelastic model
that includes finite first normal stress differences. The
continuum mechanical constitutive equation for the local stresses
can be calculated from the microscopic dumbbell model by using any
distribution function of the polymer in solution, and averaging over
the number of molecules per unit volume $n$~\cite{Bird87}:

\begin{equation}
\underline{\tau}_p+\lambda\underline{\tau}_{(1)}
=-nk_BT\lambda\underline\gamma_{(1)}, \label{Oldroyd-B}
\end{equation}
where $\lambda$ represents the time constant of the Hookean
dumbbells, $k_BT$ is the thermal energy and $\underline\gamma_{(1)}$ is
the velocity gradient tensor. In principle, one can now calculate
the temporal evolution of the stress tensor $\underline{\tau}_p$ for
any flow situation. In the following two subsections, the relevant
results for shear and elongational flow are
presented.

\subsubsection{Shear flow}

For the polymer contribution to the shear stress one finds

\begin{equation}
\tau_{p,xy}=\tau_{p,yx}=-nk_BT\lambda\dot\gamma ,\label{taup}
\end{equation}
where $\dot\gamma$ is the shear rate, and the first normal stress
difference is given by:

\begin{equation}
N_1=\tau_{p,xx}=-2nk_BT{\lambda}^2\dot\gamma^2. \label{N1}
\end{equation}

The first normal stress coefficient is defined as
$\Psi_1=\frac{N_1}{\dot\gamma^2}$, this then leads to~\cite{Bird87}:

\begin{equation}
%\Psi_1=2nk_BT\lambda_0^2
\lambda=\sqrt{\frac{\Psi_1}{2nk_B T}} \label{Psi1}.
\end{equation}

\subsubsection{Elongational flow}

In uniaxial elongation flow the polymers are stretched the most
efficiently. They uncoil in the flow and build up large elastic
stresses, and the elongational viscosity of the liquid increases
severely compared to the solvent viscosity. Assuming a cylindrical
filament in the CaBER, for a polymer solution, the elongational
viscosity is the ratio between the first normal stress difference
and the elongation rate and an analysis of the Oldrd-B model yields \cite{Schummer83,Clasen06b}

\begin{equation}
\eta_{e}(t)=\frac{\tau_{zz}-\tau_{rr}}{\dot{\epsilon}(t)}= 3 \left(\frac{\sigma}{h_{0}}\right)^{\frac{4}{3}} \left(\frac{1}{nk_BT}\right)^{\frac{1}{3}}
\lambda \exp(t/(3 \lambda)),\label{elongational_viscosity}
\end{equation}
where $\tau_{zz}$ and $\tau_{rr}$ are the normal stresses in the
stretching and the radial direction, $\dot\epsilon$ is the
elongation rate, $\sigma$ the surface tension that drives the thinning process and $h_0$ the thickness of the filament when it is formed. Here, we see
that the polymer relaxation time $\lambda$ characterizes the
exponential growth of the elongational viscosity and the normal
stress coefficient amplitude, two quantities that are experimentally
accessible. Given this, it should be possible to experimentally
relate normal stress characteristics to elongational properties of
flexible polymer solutions.

In some cases, a transition from exponential thinning behavior of the filament to a linear regime has been reported. This is often used to extract a plateau value for the elongational viscosity, but the regime is typically very small and the validity of this approach might be questionable. If, indeed, polymers are stretched to a maximum, the pinch off dynamic should be governed by self similar laws that reflect the full hydrodynamic dynamics\cite{Sattler08}.
%\subsection{Non-Hookean dumbbell models}

\subsection{FENE-P model}

One model that is supposed to describe flexible polymers reasonably
well and includes their finite extensibility is the FENE model
\cite{Bird87}. Furthermore, it can describe, in some limits, the
effect of shear thinning. In our case, the solution for a
stationary shear flow gives the following expressions for the polymer
part of the viscosity and the normal stresses, respectively
\cite{Bird87}:

\begin{equation}
\eta_p=\frac{2nk_BTb}{\dot{\gamma}(b+2)}\sqrt{\frac{b+5}{6}}
sinh\left(\frac{1}{3}arsinh\left(\frac{3\dot{\gamma}\lambda(b+2)}{2(b+5)}\sqrt{\frac{6}{b+5}}\right)\right)
\label{eta_p_FENE-P},
\end{equation}

\begin{equation}
N_1=\frac{4nk_BTb(b+5)}{3(b+2)}\left(sinh\left(\frac{1}{3}arsinh\left(\frac{3\dot{\gamma}\lambda(b+2)}{2(b+5)}\sqrt{\frac{6}{b+5}}\right)\right)\right)^2
\label{N1_FENE-P}.
\end{equation}

Here, the parameters $b$ and $\lambda$ represent the finite extensibility
and the molecular relaxation time of the polymers, respectively, and
are left as independent fit parameters. Small values of $b$ result
in a description of very stiff polymers, whereas, for the limit of
infinite $b$, the equations \ref{eta_p_FENE-P} and \ref{N1_FENE-P}
approach the respective equations for an Oldroyd-B fluid. Therefore,
the FENE-P model is able to describe the shear thinning of a given
polymer solution only as long as $b$ remains small.

%-------------------- SEC III -----------------------------
\section{Experimental methods} \label{sec3}
\subsection{Sample preparation}
%---------------------------------------------------------

Two different polymers in glycerol-water mixtures were used (see
Table \ref{table1}): Polyacrylamide (PAAm) with a molecular weight
of $5-6\times10^6$ g/mol (Fluka) and Polyethyleneoxide (PEO) with
$4\times10^6$ g/mol (Aldrich).
%and Xanthan gum with $933,75$ g/mol (Fluka).

PAAm and PEO are flexible polymers. The different solubilities of the
polymers limited us to different concentration ranges (cf. Table
\ref{table1}). For the solutions with low polymer concentrations in
low viscosity solvents, the normal stresses could not be reasonably
well determined. For the high polymer concentrations in very viscous
solvents, it was not always possible to perform CaBER measurements
because the filament showed strong deviations from a simple
exponential thinning process. Finally, some solutions with a high
polymer concentration in a glycerol-rich solvent could not be
prepared. The polymers did not dissolve and the solution remained
turbid. Aside from these limitations, we prepared and characterized all
polymers with concentrations from 75ppm to 9600ppm, doubling the
concentration from one solution to the next, and in glycerol-water
mixtures, starting with water up to 80\% glycerol in steps of 20\%.Estimations from intrinsic viscosity measurements and from data from the literature indicated that the overlap concentrations for all solutions were on the order of $c^*\approx 500ppm$.

All solutions were prepared with the following protocol: the
respective amounts of polymers were slowly poured in water first.
After some minutes of swelling at rest, the polymer-water mixture
was gently stirred for 24 hours. For the glycerol-water solutions,
the respective amount of glycerol was added and the solution was stirred
again for 24 hours. All solutions were measured within one day, in
order to minimize degradation.

\begin{table} [h]
\begin{tabular}{|l|l|l|}
  \hline
  polymer & solvent & concentration c [ppm] \\
  \hline
  PAAm (5-6Mio) & 40/60 glycerol/water & 600, 1200, 2400 \\
   & 60/40 glycerol/water & 300, 600, 1200, 2400 \\
   & 80/20 glycerol/water & 150, 300, 600, 1200, 2400 \\ \hline
  PEO (4Mio) & water & 1200, 2400, 4800 \\
   & 20/80 glycerol/water & 1200, 2400, 4800 \\
   & 40/60 glycerol/water & 150, 300, 600, 1200, 2400, 4800 \\
   & 60/40 glycerol/water & 150, 300, 600, 1200, 2400 \\ \hline
%  Xanthan & water & 2400, 4800, 9600 \\
%   & 20/80 glycerol/water & 600, 1200, 2400, 4800 \\
%   & 40/60 glycerol/water & 150, 300, 600, 1200, 2400, 4800 \\
%   & 60/40 glycerol/water & 150, 300, 600, 1200, 2400, 4800 \\
%   & 80/20 glycerol/water & 150, 300, 600, 1200, 2400 \\
  \hline
\end{tabular}
\caption{All measured solutions} \label{table1}
\end{table}

%------------------  ---------------------------------------
\subsection{Experimental Setups}
%------------------------------------------------------------------

\subsubsection{Capillary Breakup Extensional Rheometer}

Our Capillary Breakup Extensional Rheometer consists of two
circular stainless steel plates with a diameter of 2mm. A droplet of
the polymer solution was put between the two plates. The upper plate
was then moved away by a linear motor, which is controlled by
software. After separation of the plates, there was a
hemispherical fluid reservoir at each of the two plates. Between
these reservoirs a cylindrical filament formed, which thinned
exponentially in time. The diameter of the thinning filament was
measured with a high-speed camera (XS-5, IDT) with a resolution of
1280$\times$500, at a frame rate of 2100Hz. The filament was imaged
with a microscope objective (Nikon) with fourfold magnification and the diameter $h(t)$ was determined from the digital shadowgraphs (fig. \ref{CaBER-PEO}) with a threshold algorithm.

The elongation rate $\dot{\varepsilon}(t)$ in a capillary thinning
experiment is given by \cite{Bird87}:

\begin{equation}
\dot{\varepsilon}(t)=-2\frac{\partial_t h(t)}{h(t)},
\label{elongation_rate}
\end{equation}
where $h(t)$ is the minimum diameter of the cylindrical filament.
The filament thins exponentially with time according to:

\begin{equation}
h(t)=h_0\cdot e^{-t/\lambda_C}, \label{diameter_exp}
\end{equation}
where $h_0$ is the diameter at which the exponential thinning starts,
and $\lambda_C$ is a characteristic time constant of the polymer
solution . From equation \ref{diameter_exp} and
\ref{elongation_rate}, it follows that the elongation rate is
constant and given by

\begin{equation}
\dot{\varepsilon}(t)=\frac{2}{\lambda_C}.\label{constant_elongation_rate}
\end{equation}

The thinning dynamics result from the surface tension
that tends to thin the filament and elongational viscosity
($\eta_e$) that resists thinning:

\begin{equation}
\eta_e \dot\epsilon (t)=\frac{2\sigma}{h(t)}.
\end{equation}

Using equation \ref{constant_elongation_rate}, this leads to:

\begin{equation}
\eta_e(t)=\frac{\lambda_C \sigma}{h(t)}.\label{simple_elonagtional_viscosity}
\end{equation}

This means that the elongational viscosity increases exponentially
within time with the characteristic time constant $\lambda_C$, in accordance with equation \ref{elongational_viscosity} that was deduced by an analysis of the Oldroyd-B model.

We should also mention that equation
\ref{simple_elonagtional_viscosity} is only valid as long as the
filament thins exponentially\cite{Renardy94, Renardy95}.

\subsubsection{Rheometer} \label{sec4.2}

The shear flow measurements are performed in a commercial rheometer
(Haake MARS) with cone-plate geometry. Here, we used a cone with
an angle of 2$^\circ$; the cone and the plate are 60mm in diameter,
and the probe volume is 2.0ml. The cell is tempered to 20$^\circ$C
via a Haake PhoenixII with an accuracy of 0.01$^\circ$C.

Data points were taken at shear rates between 1$\frac{1}{s}$ and
2000$\frac{1}{s}$. The shear rate has been changed on a logarithmic
scale between $1 \frac{1}{s}$ and 250$\frac{1}{s}$, and in linear
steps of 25$\frac{1}{s}$ for higher shear rates. Please note that the typical elongational rates that have been observed in CaBER measurements were of the same order of magnitude.

%--------------------------- SEC V ------------------------
\section{Measurements and Results}
\label{sec5}
%-----------------------------------------------------------

\subsection{Capillary Breakup Extensional Rheometer}

\begin{figure} [h!]
\begin{center}
\includegraphics [angle=0, width=0.65\columnwidth]{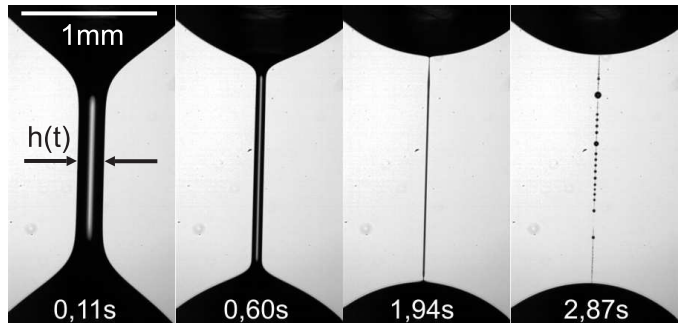}
%\vspace{0.5cm}
\caption{Image sequence of capillary breakup of a PEO filament:
the first two pictures show the cylindrical filament thinning
exponentially with time; the third picture is taken directly before
the onset of the beads-on-a-string structure and the fourth one is
the last one before final breakup (with completely formed
beads-on-a-string structure). The time scale, which is depicted at
the bottom of each image, has its starting point (0s) at the
beginning of the measurement. The time between the first and the
last picture in this sequence is about 2.76s. The beads-on-a-string
structure has already been described in a detailed way by several
authors~\cite{Rodd05,Oliveira05,Oliveira06,Clasen06b,Sattler08}.}\label{CaBER-PEO}
\end{center}
\end{figure}

Our CaBER measurements are carried out by putting a droplet of the
liquid that we want to investigate on the lower, fixed plate,
and bringing the upper, movable plate into contact with the droplet.
Then, the upper plate is pulled upwards and, thus, there is a thinning
liquid bridge between the two plates (see Figure \ref{CaBER-PEO}).

During this capillary thinning there are two different regimes: In
the first one, the filament thins exponentially with time (cf.
Equation \ref{diameter_exp}). Typically, one also observes a second
regime that is reached after the exponential one when the polymers
are fully stretched. In this regime, the sample fluid might be seen
as a Newtonian fluid but with an increased elongational viscosity.
The linear behavior is consistent with theoretical and experimental
considerations on Newtonian liquids. However, there exists no
consistent approach to describe these two regimes by use of the same
material parameters. Here, we restrict ourselves to the first regime,
where the relaxation time $\lambda_C$ is extracted (cf. Equation
\ref{diameter_exp}).

\begin{figure} [h!]
\begin{center}
\includegraphics [angle=0, width=0.88\columnwidth]{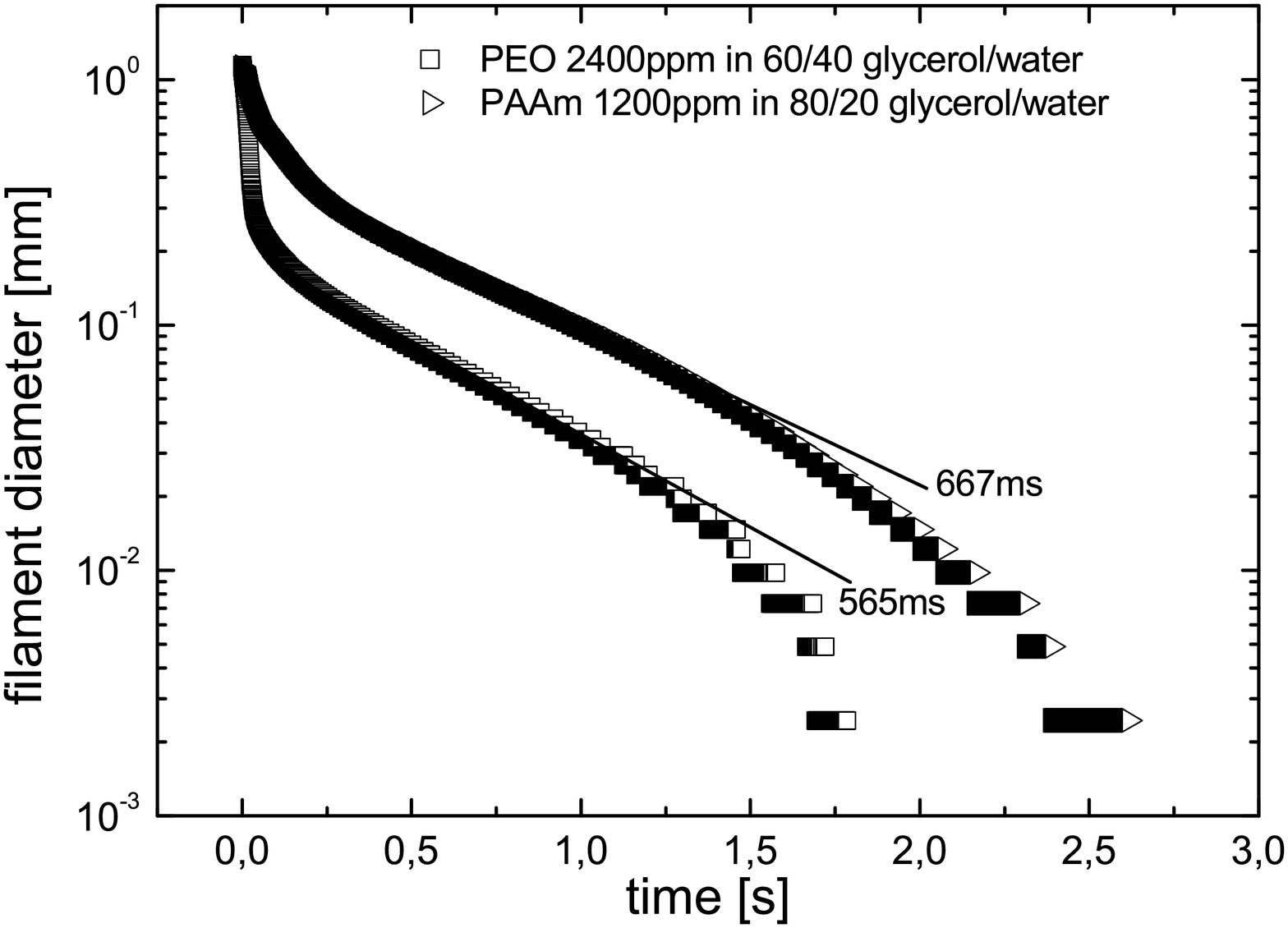}
%\vspace{0.5cm}
\caption{CaBER-measurements of two different polymers (high
concentration and high solvent viscosity). The straight lines are
the respective exponential fits to the data from which the
relaxation times are determined. %It can be seen that the relaxation
%times vary significantly between the flexible polymers and the stiff
%one.
}\label{CaBER-Messungen}
\end{center}
\end{figure}

\subsection{Rheometer}

The rheometric measurements have been carried out in a straight forward manner
with the procedure mentioned in section \ref{sec4.2}. Inertial effects
occurring at high shear rates are corrected for \cite{Macosko94}.

We could then deduce the polymer relaxation time $\lambda_N$ from normal
stress measurements using equation~\ref{Psi1}.
Furthermore, the respective fits for the FENE-P model were performed
according to equation \ref{eta_p_FENE-P} and equation
\ref{N1_FENE-P}.

\begin{figure} [h!]
\begin{center}
\includegraphics [angle=0, width=0.88\columnwidth]{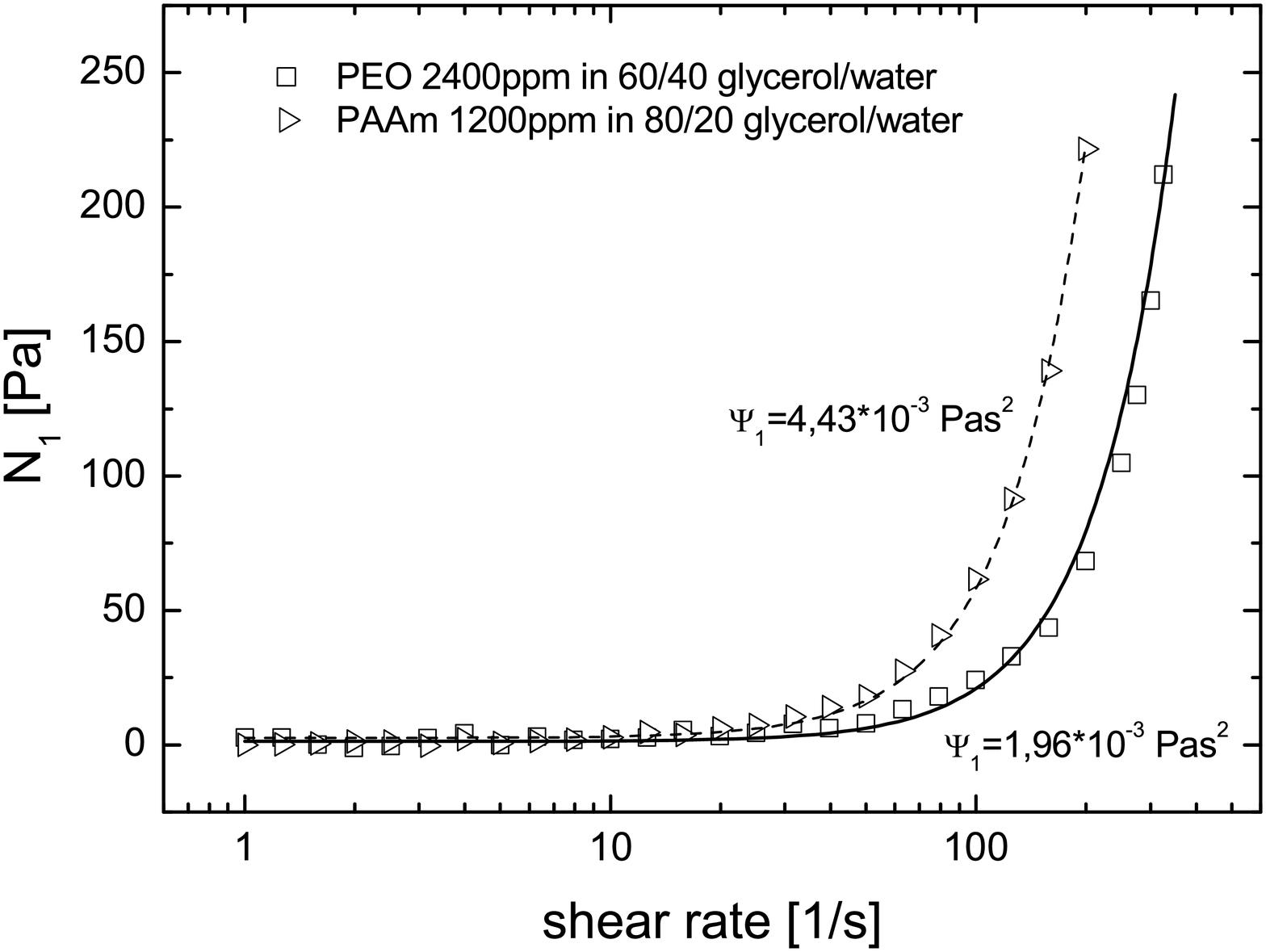}
%\vspace{0.5cm}
\caption{Rheometric measurements of two different polymer types at
high concentration and high solvent viscosity. The lines
%for PAAm and PEO
 are the respective quadratic fits.}\label{Rheo-Messungen}
\end{center}
\end{figure}

Figure \ref{Rheo-Messungen} shows two exemplary data sets for the
first normal stress difference of the different types of polymers.
The data sets are fitted with a quadratic power law (Oldroyd-B fluid).

\begin{figure} [h!]
\begin{center}
\includegraphics [angle=0, width=0.49\columnwidth]{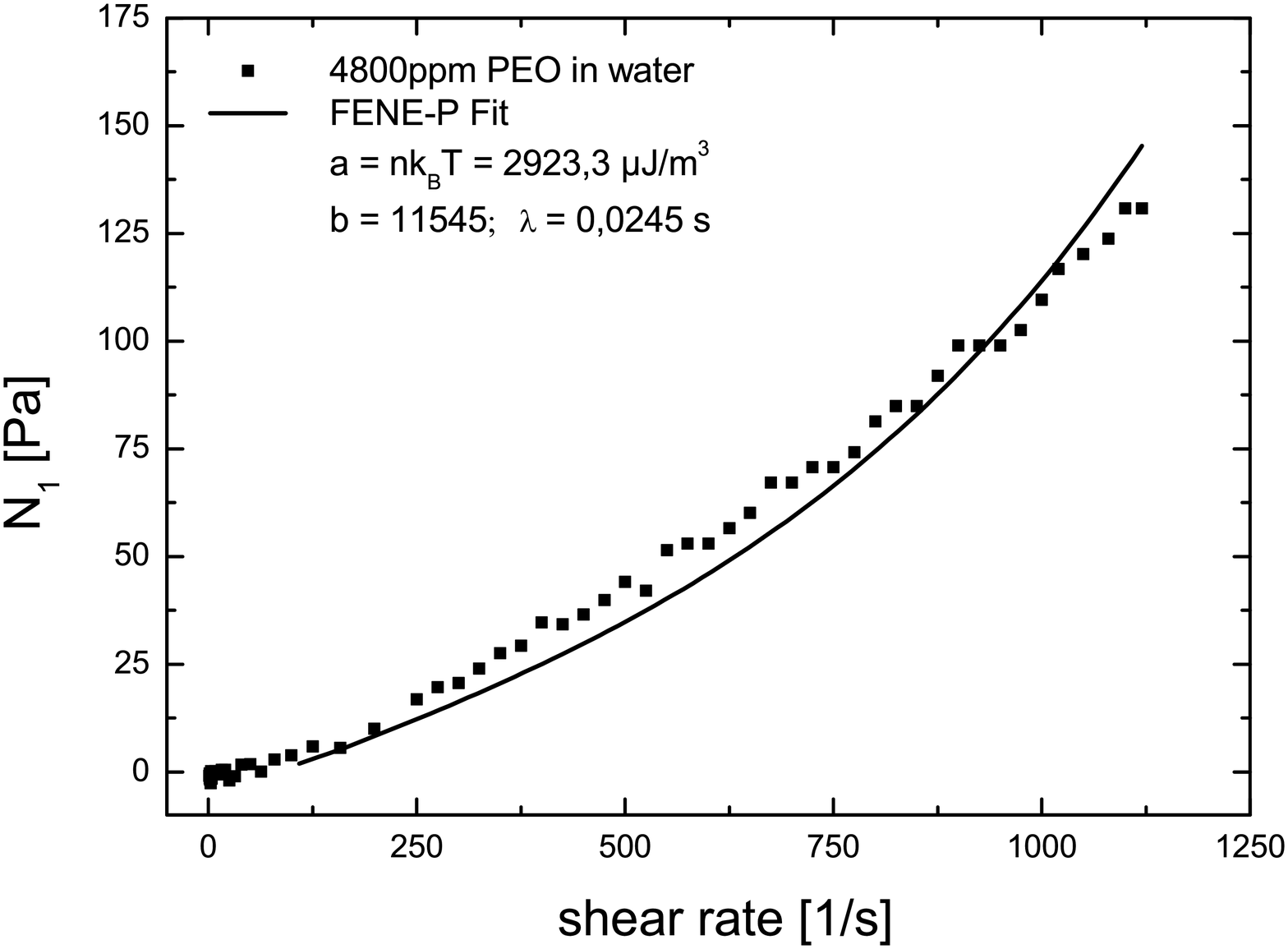}
\includegraphics [angle=0, width=0.49\columnwidth]{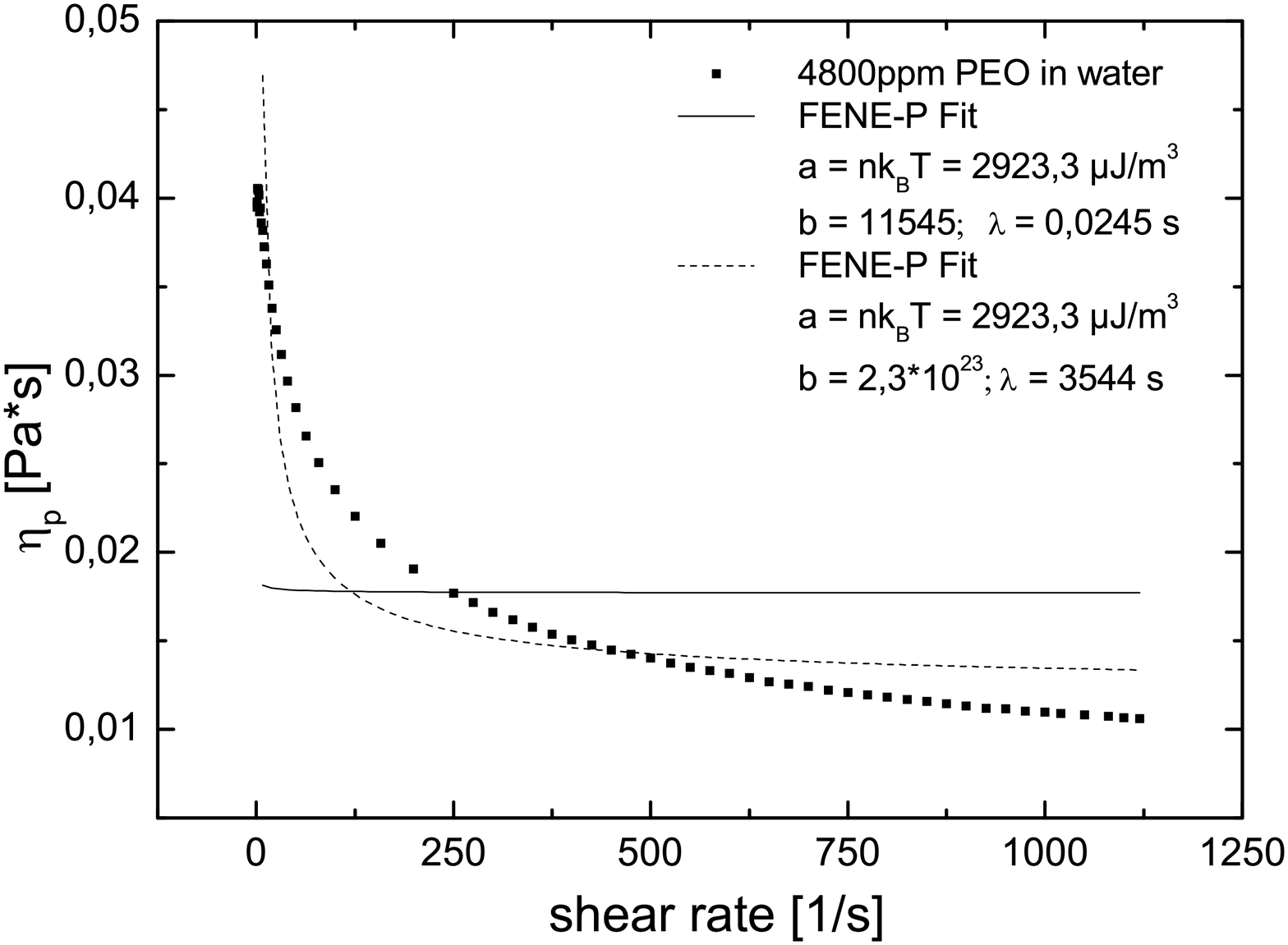}
%\vspace{0.5cm}
\caption{Left: FENE-P fit for the first normal stress difference of
4800ppm PEO dissolved in water. Respective parameters are shown in the
legend and are used later to fit the polymer viscosity.\newline
Right: Shear thinning effect of 4800ppm PEO in water. Respective
fits are done with the FENE-P model, and the parameters determined from
normal stress measurements (straight line) as well as the FENE-P model
and free parameters (dotted line).} \label{FENE-P}
\end{center}
\end{figure}

Finally, we would like to discuss the FENE-P fits for the PEO
solutions. An analysis of the normal stress data yielded
relaxation times similar to those of the elastic dumbbell model, but with a relatively
large parameter b within the FENE model (Figure \ref{FENE-P}),
meaning a high flexibility.

However, the same set of parameters does not describe the shear
thinning of the polymer part of the viscosity fit (Figure
\ref{FENE-P}).

The same result holds if the fits for $N_1$ and $\eta_p$ are
performed simultaneously. The fits nicely reproduce the normal
stress data but fail to describe the viscosity. A fit for the
viscosity data alone can be performed reasonably well, but the
parameters are far from being realistic.

Actually, the typical approach to extract a timescale from $N_1$ data is, indeed, to combine equation \ref{taup} and \ref{N1}, and to divide the normal stress coefficient $\Psi_1$ by the polymeric part of the viscosity. In this way the number density drops out of the equation but the drawback is that due to the shear thinning of the polymeric part of the viscosity it yields to a shear rate dependent time constant, which is also called a shear thinning relaxation time. Even if the relaxation times that we extracted by this method were on the same order of magnitude as the ones that we determined by use of the microscopic quantities, the data was scattered much more strongly, mostly due to the difficulty in determining the polymeric part of the viscosity precisely. Furthermore, it is not clear which shear rate should have been chosen for comparisons with the CaBER measurements.

\subsection{Comparison between CaBER and Rheometer}

\begin{figure} [h!]
\begin{center}
\includegraphics [angle=0, width=0.7\columnwidth]{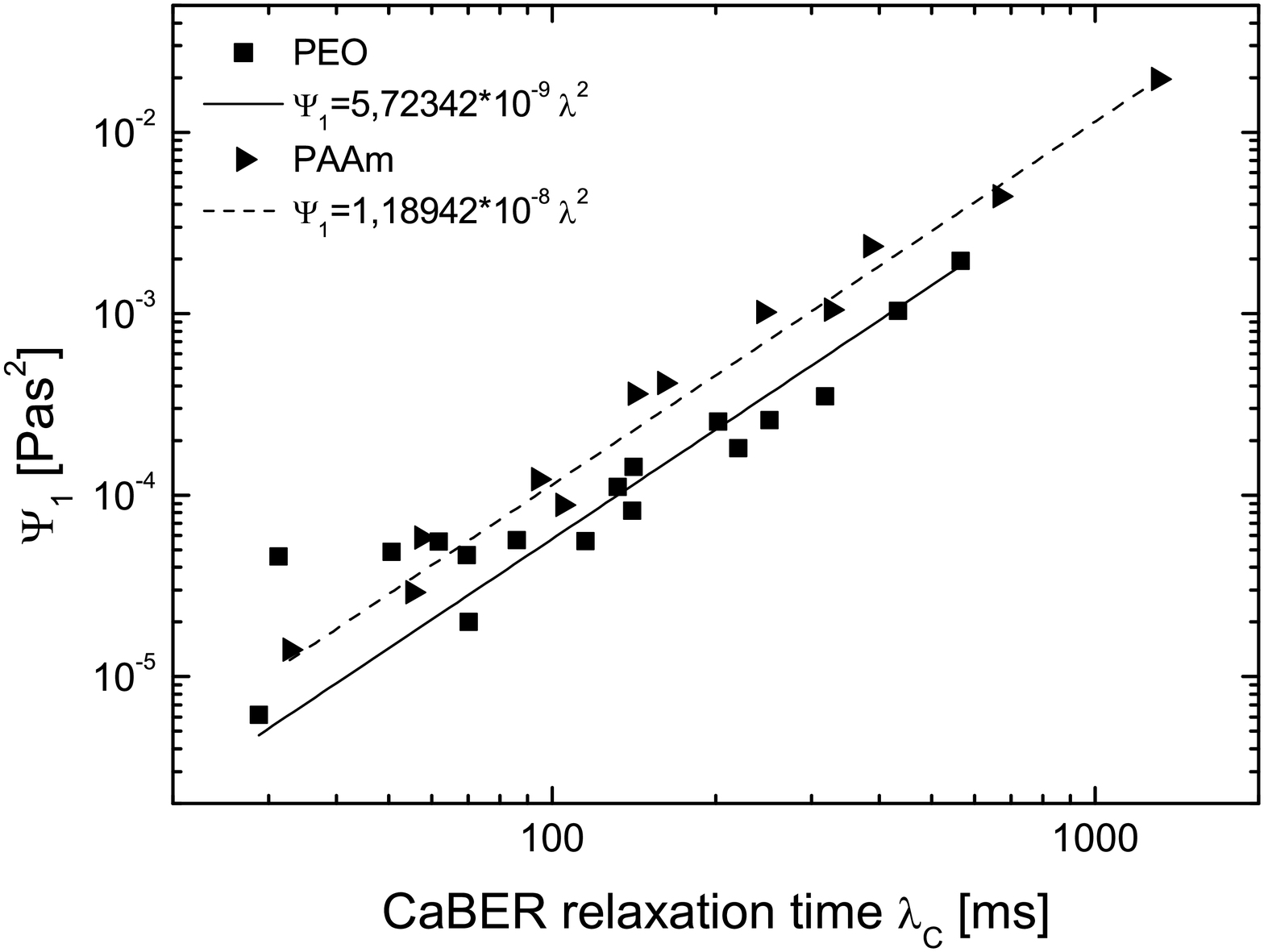}
%\vspace{0.5cm}
\caption{First normal stress coefficient as a function of the CaBER
relaxation time. Quadratic fits are applied for different polymer
types.}\label{Psi1_vs_lambda_CaBER_gesamt}
\end{center}
\end{figure}

Figure~\ref{Psi1_vs_lambda_CaBER_gesamt} shows the measured normal
stress coefficient $\Psi_1$ as a function of the characteristic time
$\lambda_C$ of filament thinning from the CaBER experiment, and
this is shown for the two types of polymer solutions (PEO and PAAm)
and for the whole range of concentrations indicated in
table~\ref{table1}. One can notice that $\Psi_1$ increases
quadratically as a function of $\lambda_C$. This indicates a
quantitative correlation between normal stresses and elongational
viscosity. This is what we expected as the microscopic origin of the
two effects is the resistance against stretching of the flexible
chains of polymers. At first, the quadratic dependence of $\Psi_1$
on $\lambda_C$ seems to fulfill equation~\ref{Psi1}. However,
contrarily to what is expected from equation~\ref{Psi1}, the
relation between the first normal stress coefficient and the CaBER
relaxation time is found to NOT depend on polymer number density
$n$. To investigate the reasons for this discrepancy, we will
determine in the following how $\Psi_1 $, on the one hand, and
$\lambda_C$, on the other hand, scale with polymer number density $n$
and with the solvent viscosity $\eta_s$.

We would now like to deduce a functional dependence of the normal
stress and elongational viscosity data. We will use the relaxation
time $\lambda_C$ from the CaBER measurements, and the relaxation time
$\lambda_N$ and the first normal stress coefficient $\Psi_1$ from
the shear experiments.

It is already known from the literature that $\lambda_C$ strongly
depends on both polymer concentration and solvent viscosity;
$\lambda_C$ increases with increasing concentration or increasing
viscosity (cf.~\cite{Clasen06} or~\cite{Tirtaatmadja06} for
example). For $\lambda_N$, the situation is less obvious. The
Oldroyd-B model predicts that the relaxation time depends on the
solvent viscosity but not on the polymer concentration. This
directly follows from the definition of the relaxation time as
$\lambda=\frac{\zeta}{4H}$, where $\zeta$ is a friction coefficient
which linearly depends on the solvent viscosity and $H$ is the
spring constant of the Hookean dumbbell~\cite{Bird87}. Nevertheless,
experimentally determined relaxation times can  show different
behaviors. Of course, it is important if the concentration is below
or above the so called overlap concentration $c^*$. The overlap
concentration is the concentration at which two polymers in a
solution interact with each other; this can lead to, e.g., entanglements
that increase the relaxation time of the solution. Above this
overlap concentration a solution is called semidilute, whereas below
$c^*$, it is a dilute one~\cite{Rubinstein03}. The effect of polymer
interactions is not considered in the Oldroyd-B model. Still, within the limits of a continuum mechanical approach it is often possible to deduce a physical understanding of the physical origin of certain viscoelastic flow phenomena.

>From the first normal stress difference measurements we deduce the
relaxation time $\lambda_N$ from equation~\ref{Psi1} for a range of
concentrations and solvent viscosities. We can then determine scaling
laws for $\lambda_N$ in the form:
\begin{equation}
\lambda_N\propto n^{a_N}\eta_s^{b_N}\label{lambda_N_exponents_def}
\end{equation}

This corresponds to a scaling for $\Psi_1$ in the form:
\begin{equation}
\Psi_1\propto n^{(2a_N+1)}\eta_s^{2b_N}.\label{Psi1_exponents}
\end{equation}

\begin{figure} [h!]
\begin{center}
\includegraphics [angle=0, width=0.49\columnwidth]{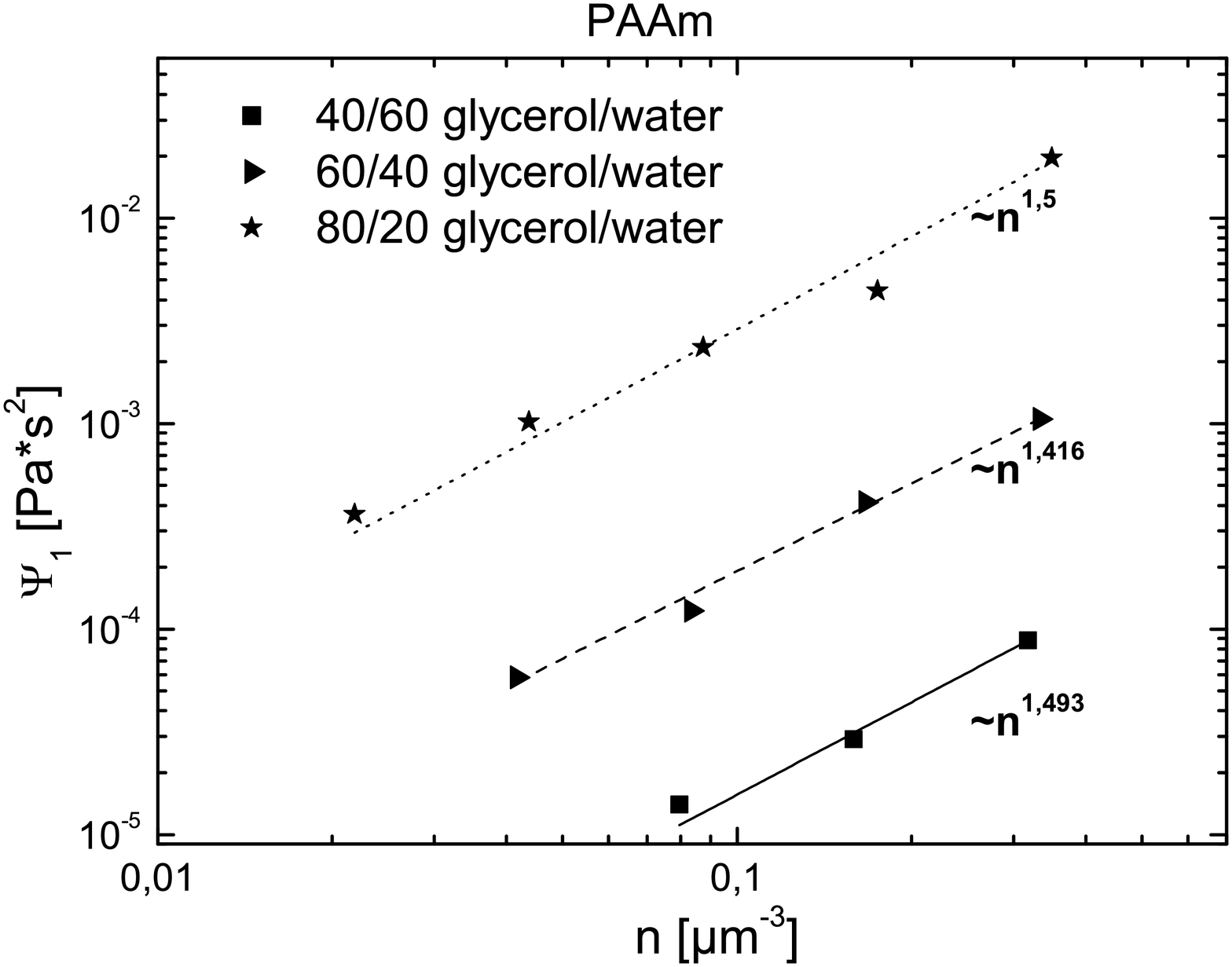}
\includegraphics [angle=0, width=0.49\columnwidth]{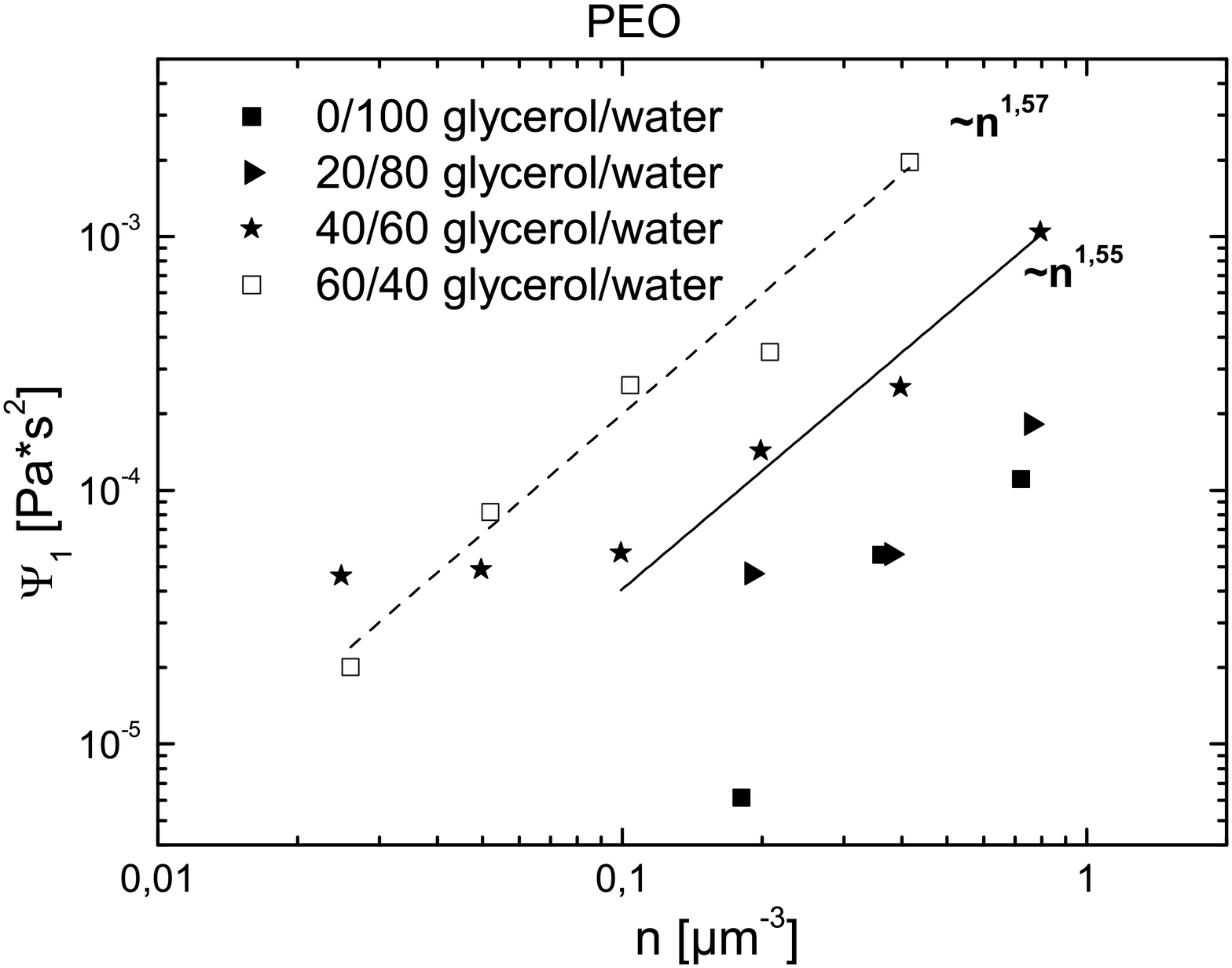}
\includegraphics [angle=0, width=0.49\columnwidth]{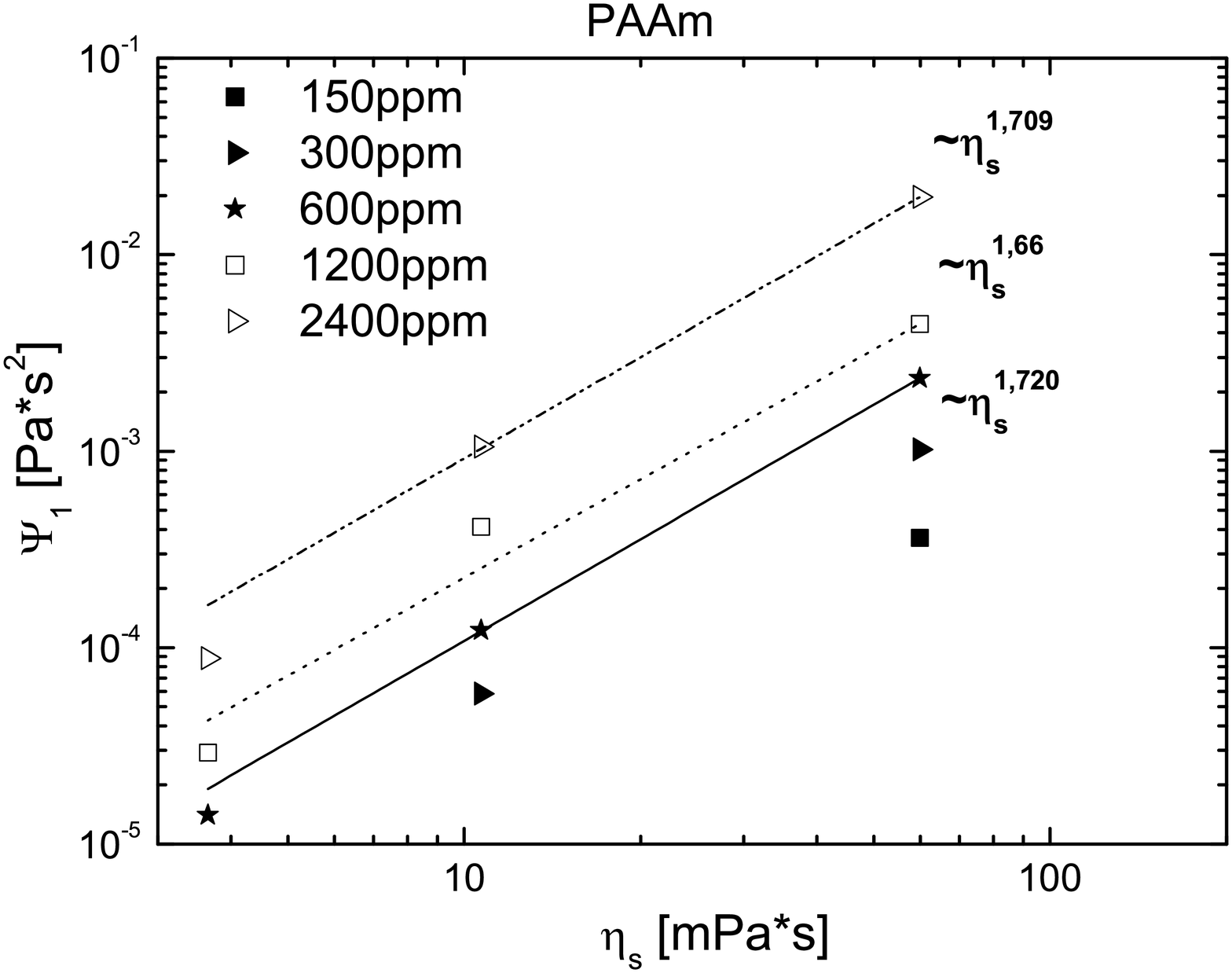}
\includegraphics [angle=0, width=0.49\columnwidth]{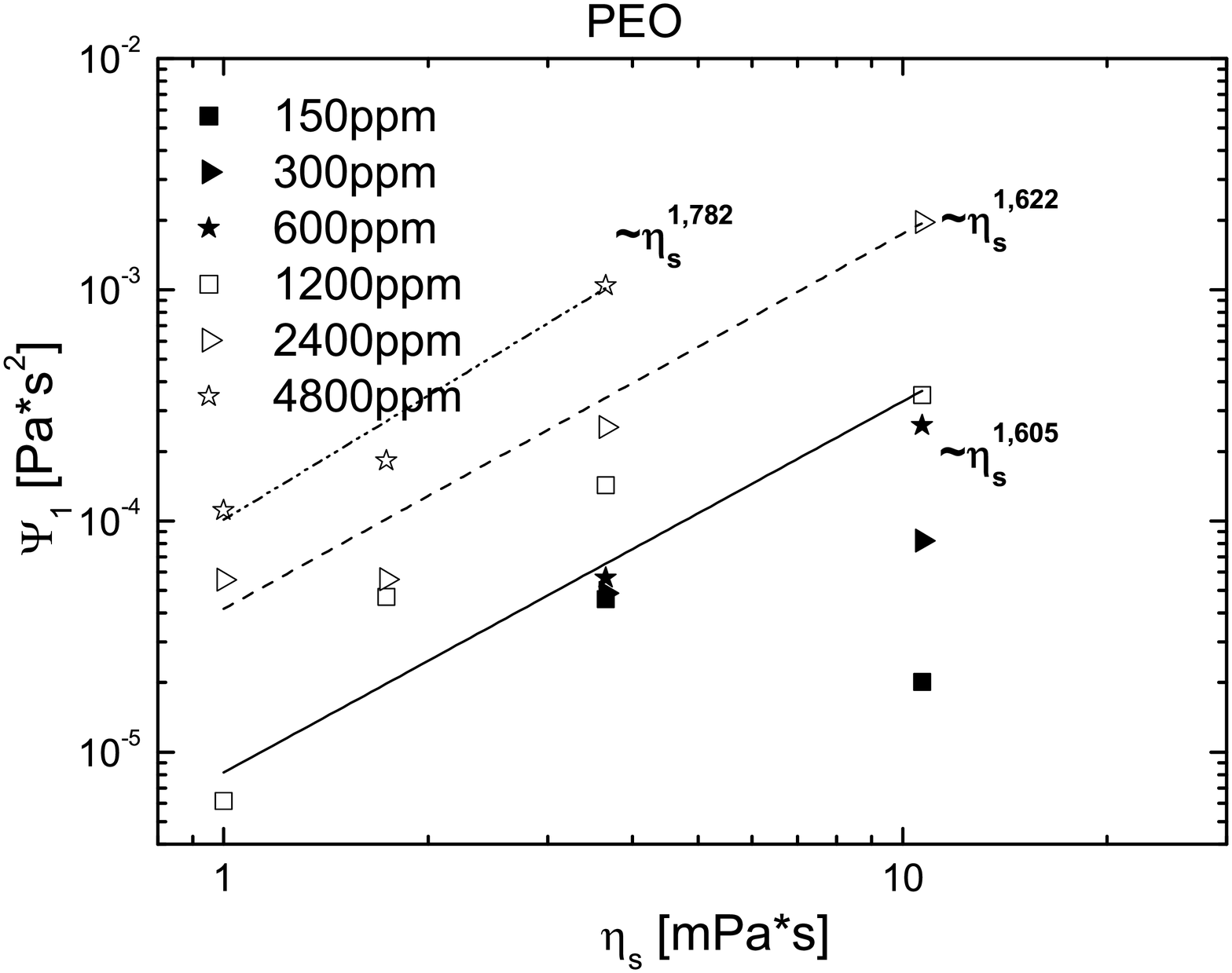}
%\vspace{0.5cm}
\caption{Upper images: First normal stress coefficient of the PAAm
and PEO solution as a function of the number density per unit
volume. Power law fits are applied for different solvent
viscosities. For the PEO, reasonable fits were only possible for the
two more viscous solutions because of the limited experimental
resolution.\newline Lower images: First normal stress coefficient of
PAAm and PEO as a function of the solvent viscosity. Power law fits
are applied for different polymer concentrations.} \label{Psi1_vs_n}
\end{center}
\end{figure}

Figure \ref{Psi1_vs_n} shows the dependency of $\Psi_1$ on the polymer
number density $n$ and on the solvent viscosity $\eta_S$ and is well
described by power laws. $\Psi_1$ is found to increase with the polymer
concentration with an exponent around $1.5$, and with the solvent
viscosity with an exponent around $1.7$.

The averaged values for each polymer are given in
table \ref{table2}.

With the experimental values of the exponents of $\Psi_1(n)$ and
$\Psi_1(\eta_s)$ we can deduce the scaling exponents for $\lambda_N$
as follows:

\begin{equation}
\lambda_N\propto n^{a_N}\eta_s^{b_N}\approx n^{0.25}\eta_s^{0.85}.\label{lambda_N_exponents}
\end{equation}

The dependency of $\lambda_C$ on the polymer concentration $n$ and
on the solvent viscosity $\eta_s$ was already investigated by
Amarouchene et al.~\cite{Amarouchene01}, Clasen et
al.~\cite{Clasen06}, Tirtaatmadja et al.~\cite{Tirtaatmadja06} and
Rodd et al.~\cite{Rodd05}. Again, we look for scaling
laws in the form:

\begin{equation}
\lambda_C\propto n^{a_C}\eta_s^{b_C}%\approx n^{0.8}\eta_s.
\label{lambda_C_exponents}
\end{equation}

\begin{figure} [h!]
\begin{center}
\includegraphics [angle=0, width=0.49\columnwidth]{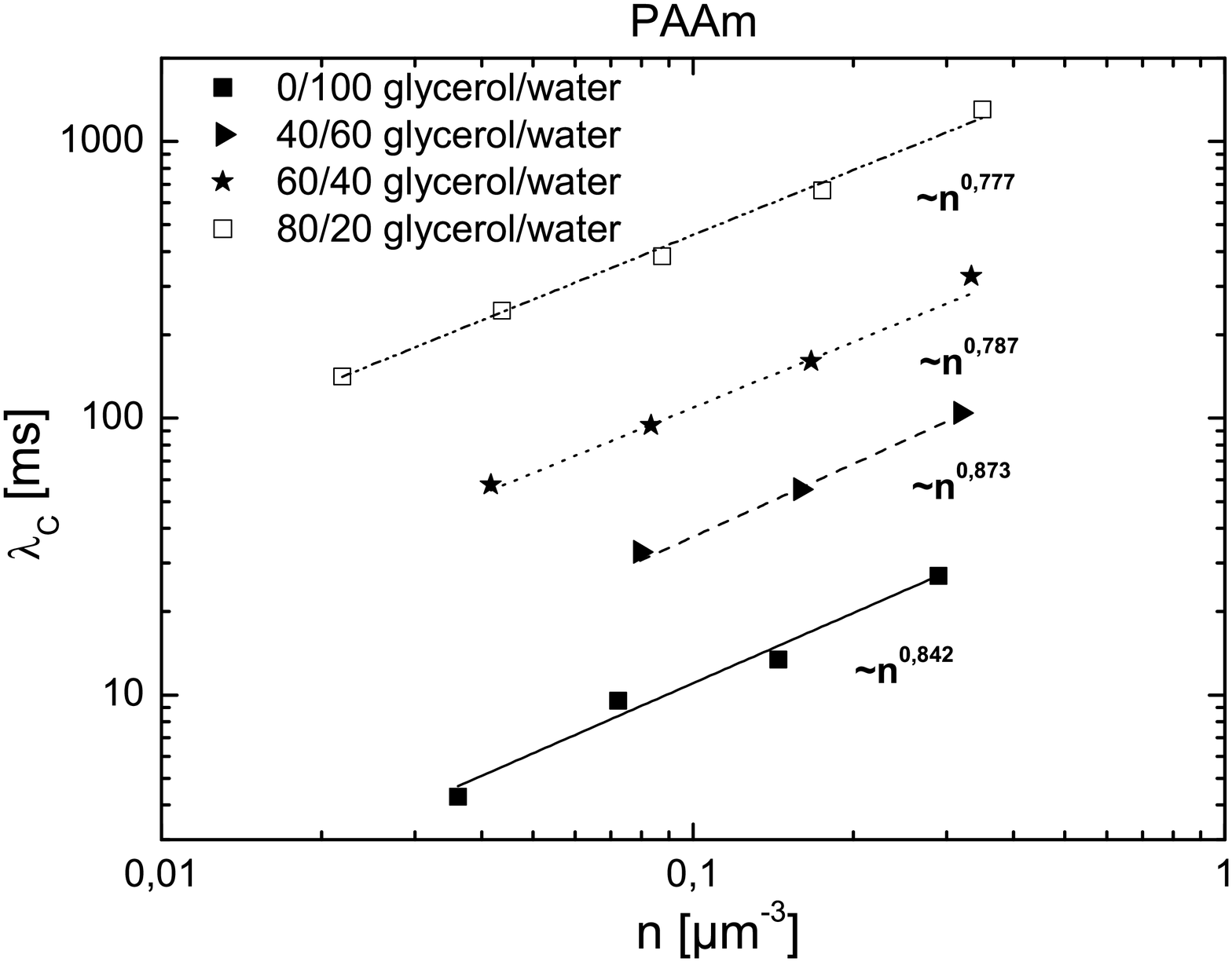}
\includegraphics [angle=0, width=0.49\columnwidth]{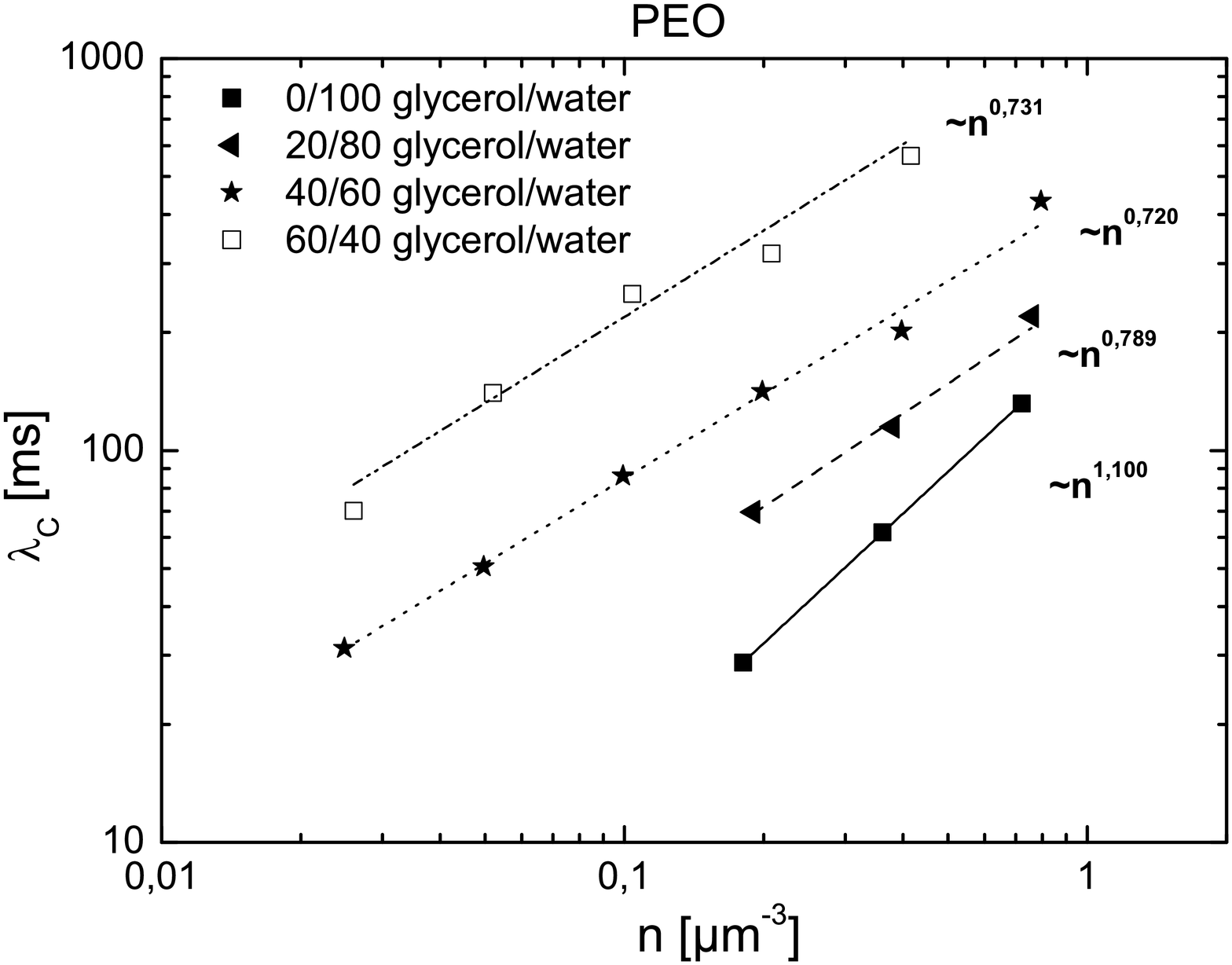}
\includegraphics [angle=0, width=0.49\columnwidth]{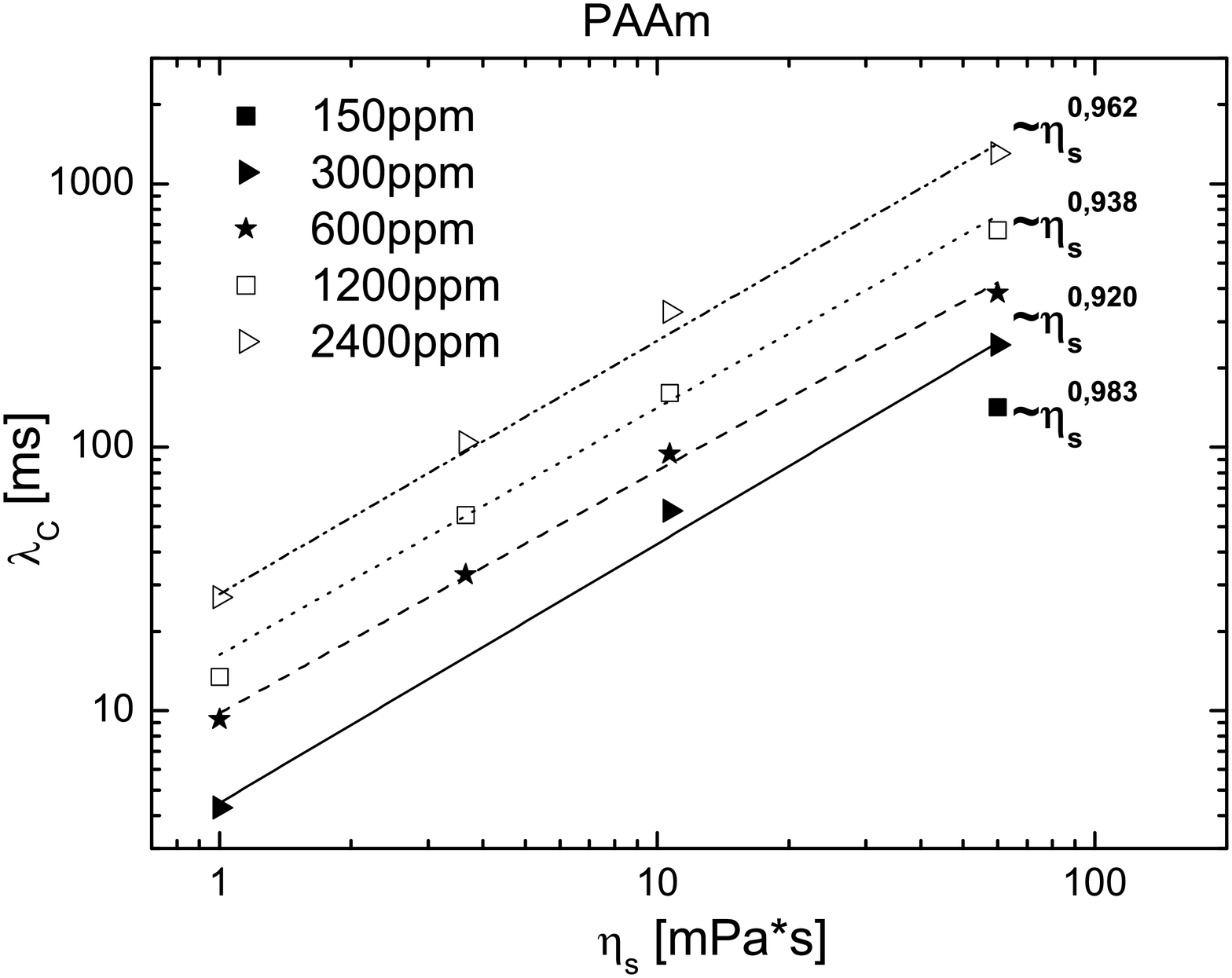}
\includegraphics [angle=0, width=0.49\columnwidth]{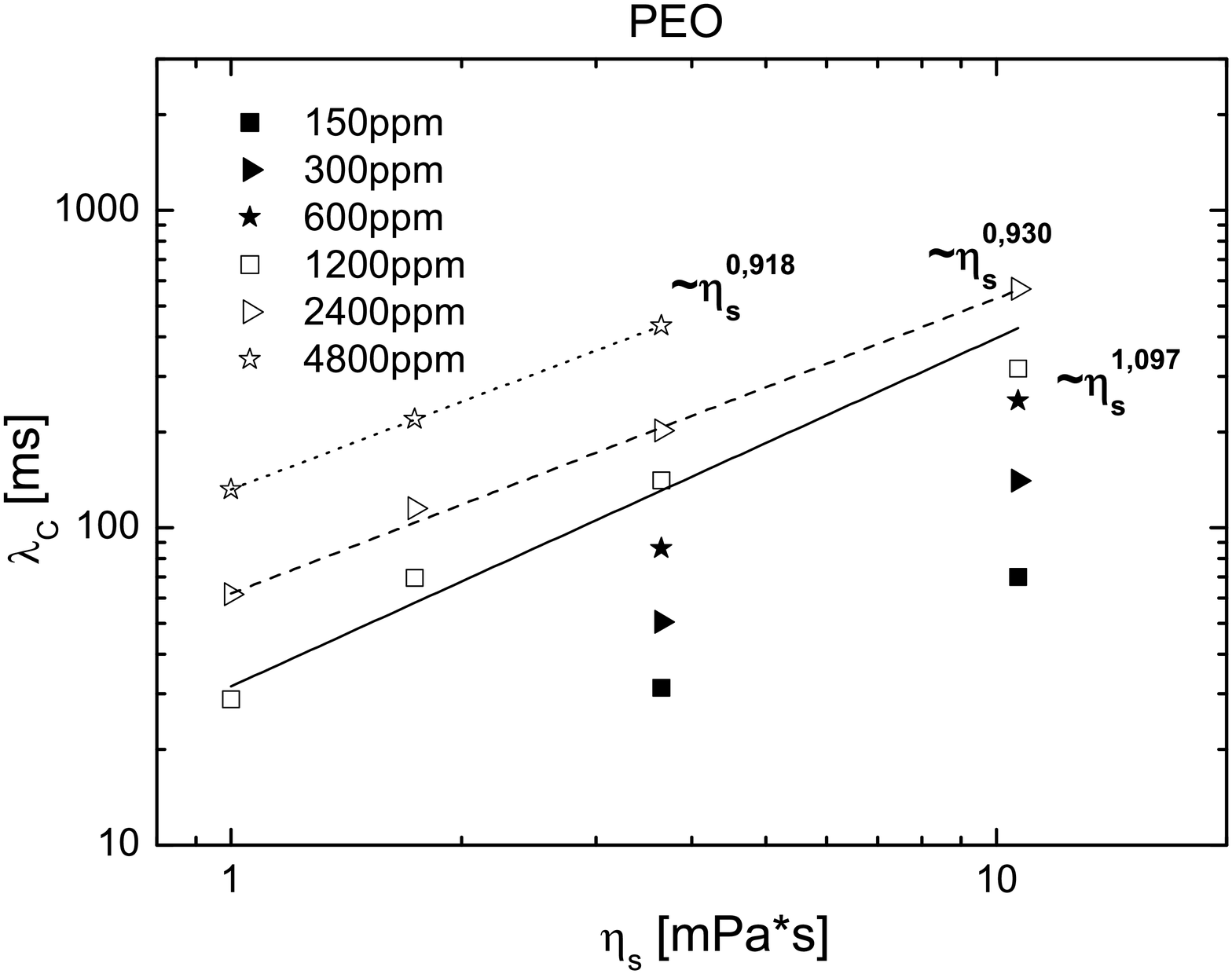}
%\vspace{0.5cm}
\caption{Upper images: CaBER relaxation time of PAAm and PEO as a
function of the number density per unit volume. Power law fits are
applied for different solvent viscosities.\newline Lower images:
CaBER relaxation time of PAAm and PEO as a function of the solvent
viscosity. Power law fits are applied for different polymer
concentrations.} \label{lambda_vs_n}
\end{center}
\end{figure}

Figure \ref{lambda_vs_n} shows that the relaxation time deduced for
filament thinning is well described by power laws. We found that
$\lambda_C$ increases linearly with solvent viscosity and with an
exponent of $0.8$ on the polymer number density. The averaged
values of the exponents are tabulated in table \ref{table2}. Even if we are mostly interested in scaling laws, we should mention that the absolute values of $\lambda_N$ and $\lambda_C$ might differ by more than one order of magnitude with, typically, $\lambda_N > \lambda_C$ and only at higher concentrations $\lambda_N \propto \lambda_C$ or even $\lambda_N < \lambda_C$.

 For PEO
in glycerol/water mixtures, Clasen et al.~\cite{Clasen06} and
Tirtaatmadja et al.~\cite{Tirtaatmadja06} both found that the
relaxation time measured in an elongation experiment varies with
$n^{0.65}$, whereas Amarouchene et al.~\cite{Amarouchene01} observed
an exponent of about 0.82 for PEO in water with a droplet
detachment experiment.

A theoretical prediction for the dependence of the polymer relaxation time on the concentration for disentangled semidilute polymer solutions is given by Rubinstein and coworkers Ref.\cite{Rubinstein03}. The exponent depends on the solvent quality and might vary from $0.31$, a good solvent, to $1$ a $\theta$ solvent. In this sense, one could simply explain the differences of $a_C$ and $a_N$ by a change in solvent quality. Even if phase separation effects in elongational flow have been reported \cite{Sattler08}, this does not seem to be likely - in both experiments the solutions were the very same. Similarly, one could imagine the dynamic in the CaBER experiment is better described by a theory of entangled semidilute solutions due to a larger elongation of the polymers and a larger effective volume. But, in this case, the exponents should be well larger than $1$ \cite{Rubinstein03}, in contradiction to our observations. Clasen et al. ~\cite{Clasen06} also compared their findings for $a_C$ with the theoretical predictions by Rubinstein. They interpret a strong dependence of $\lambda_C$, even in the dilute regime, by an effective overlap concentration that is much smaller in elongational flow than in shear flow. However, Amarouchene et al.\cite{Amarouchene01} found a dependency of $\lambda_N$ down to 10 ppm.

\begin{table} [h]
\begin{center}
\begin{tabular}{|c|c|c|}
  \hline
   & PAAm($M_w$=5-6Mio) & PEO($M_w$=4Mio) \\
  \hline
  $2a_N+1$ & \raisebox{-1.5ex}[1.5ex]{1,470$\pm$0,038} & \raisebox{-1.5ex}[1.5ex]{1,560$\pm$0,010} \\
  ($\Psi_1$ vs. $n$) &  &  \\ \hline
  $2b_N$ & \raisebox{-1.5ex}[1.5ex]{1,696$\pm$0,026} & \raisebox{-1.5ex}[1.5ex]{1,670$\pm$0,080} \\
  ($\Psi_1$ vs. $\eta_s$) &  &  \\ \hline
  $a_C$ & \raisebox{-1.5ex}[1.5ex]{0,820$\pm$0,039} & \raisebox{-1.5ex}[1.5ex]{0,835$\pm$0,155} \\
  ($\lambda_C$ vs. $n$) &  &  \\ \hline
  $b_C$ & \raisebox{-1.5ex}[1.5ex]{0,951$\pm$0,024} & \raisebox{-1.5ex}[1.5ex]{0,982$\pm$0,082} \\
  ($\lambda_C$ vs. $\eta_s$) &  &  \\
  \hline
\end{tabular}
\caption{Averaged exponents} \label{table2}
\end{center}
\end{table}

\begin{figure} [h!]
\begin{center}
\includegraphics [angle=0, width=0.49\columnwidth]{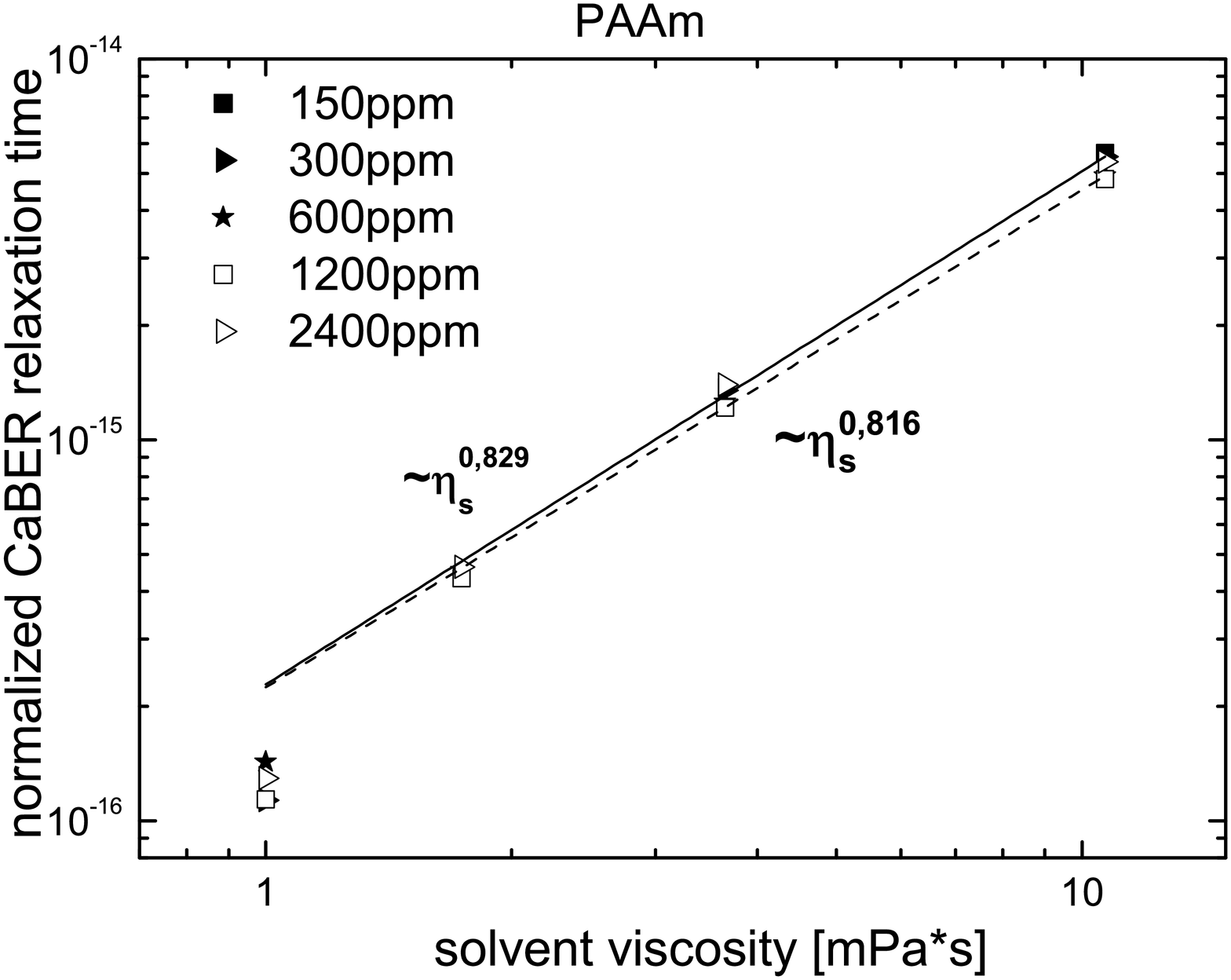}
\includegraphics [angle=0, width=0.49\columnwidth]{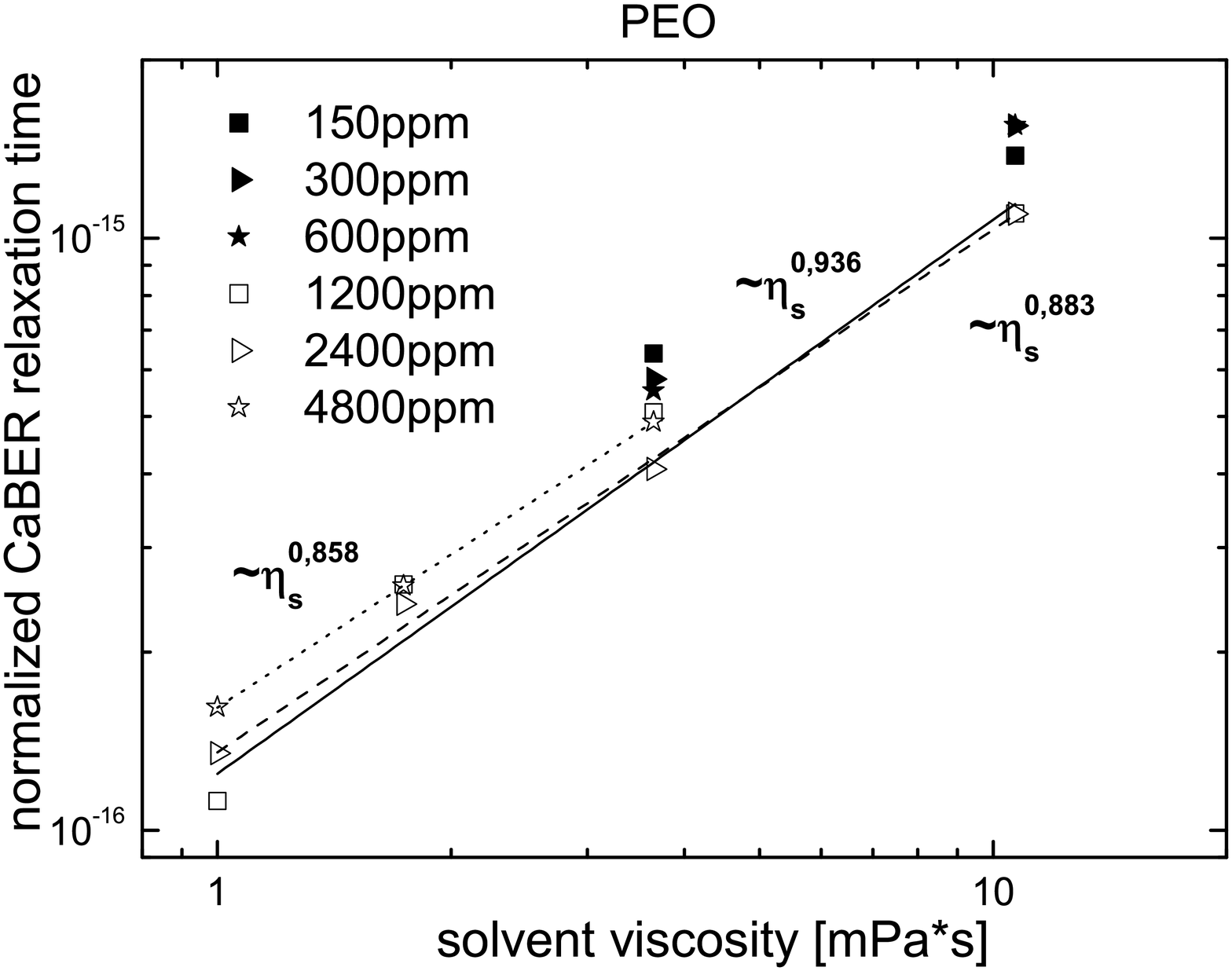}
%\vspace{0.5cm}
\caption{Rescaled CaBER relaxation times of PAAm
($\lambda_C/n^{0,820}$) and PEO ($\lambda_C/n^{0,835}$) as a
function of the solvent viscosity. Power law fits are applied for
different polymer concentrations.}\label{lambda_C_normiert}
\end{center}
\end{figure}

\begin{figure} [h!]
\begin{center}
\includegraphics [angle=0, width=0.49\columnwidth]{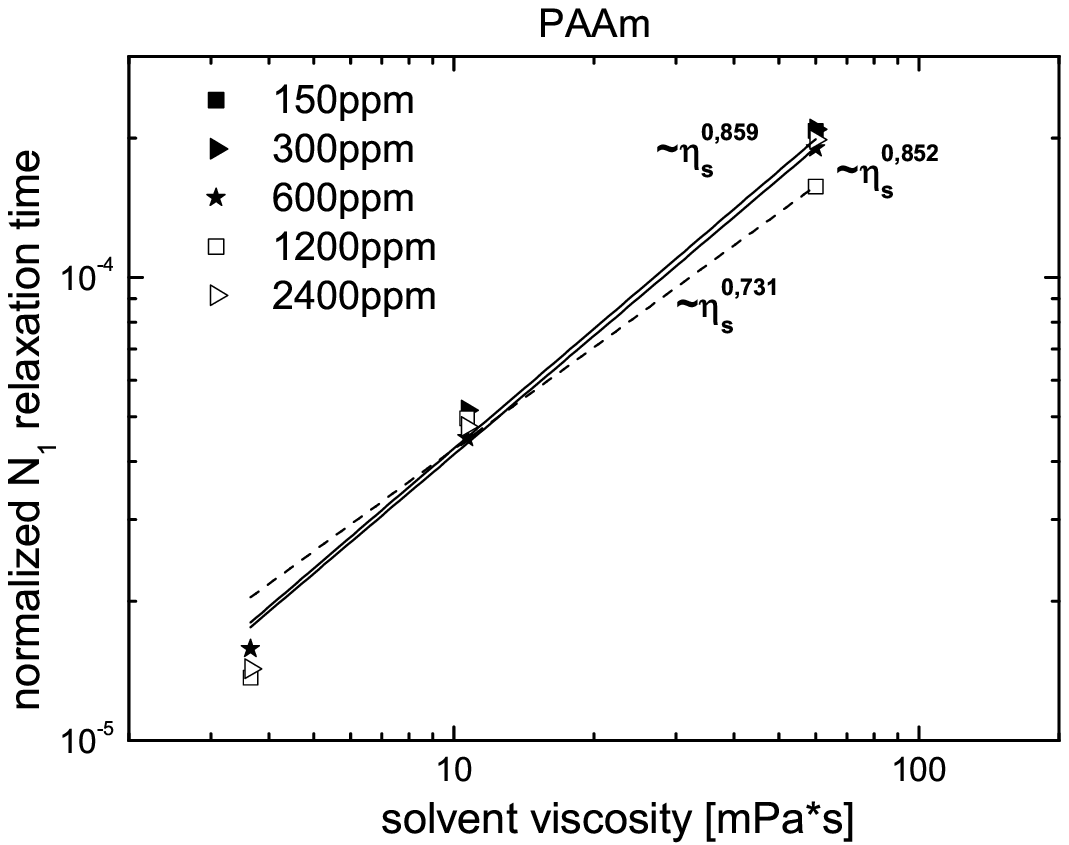}
\includegraphics [angle=0, width=0.49\columnwidth]{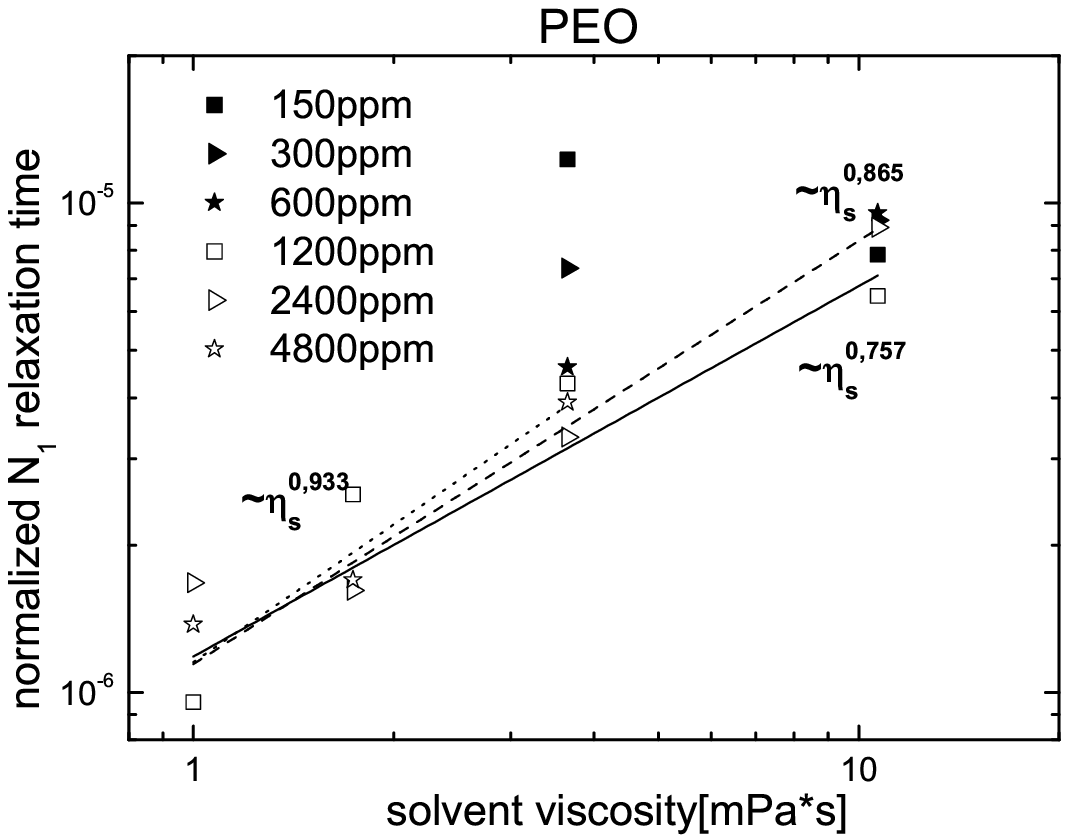}
%\vspace{0.5cm}
\caption{Rescaled normal stress relaxation times of PAAm
($\lambda_N/n^{0,235}$) and PEO ($\lambda_N/n^{0,280}$) as a
function of the solvent viscosity. Power law fits are applied for
different polymer concentrations.}\label{lambda_N_normiert}
\end{center}
\end{figure}

The CaBER relaxation time, as well as the normal stress relaxation
time, of the different polymer solutions depends on the viscosity. Thus, for
solvents with a similar exponent, these quantities can be rescaled to
coalesce on one graph. The results for both polymer types are shown
in figure \ref{lambda_C_normiert} for the CaBER and in figure
\ref{lambda_N_normiert} for the normal stress relaxation times,
respectively.

Here, one can clearly see that, for each polymer type, the rescaled
CaBER and normal stress relaxation times as a function of the solvent
viscosity fall on one curve. Apparently, the remaining dependency on
the solvent viscosity shows almost the same exponent for the CaBER
and the normal stress relaxation times, independent of the polymer type.

>From the $2a_N+1$ and $2b_N$, we are now able to compare the two
different relaxation times that were determined in the CaBER
($\lambda_C$) and in the rheometer ($\lambda_N$), respectively.
Dividing $\lambda_C$ by $\lambda_N$, we get:

\begin{equation}
\frac{\lambda_C}{\lambda_N}\propto n^{(a_C-a_N)}\eta_s^{(b_C-b_N)}\approx n^{0.5}.
\end{equation}

\begin{figure} [h!]
\begin{center}
\includegraphics [angle=0, width=0.7\columnwidth]{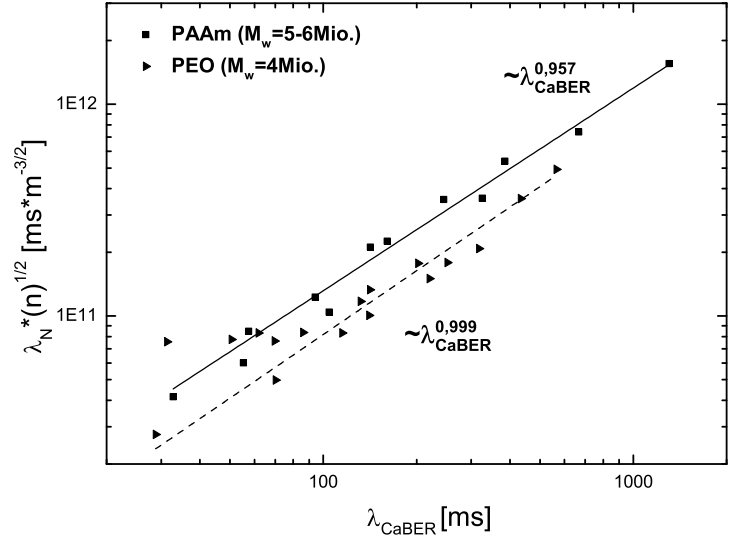}
%\vspace{0.5cm}
\caption{Rescaled normal stress relaxation time of PAAm and PEO as a
function of the CaBER relaxation time. Power law fits are applied
for different polymer
types.}\label{lambda_NWURZEL(n)_vs_Lambda_CaBER}
\end{center}
\end{figure}

In another step, one of the two relaxation times can be rescaled by
the number density per unit volume of the polymers in solution, to
adapt the normal stress relaxation time to the CaBER one. Comparing the
two dependencies of these timescales, one recognizes a difference of
$0.5$ in their potential behavior. Scaling the normal stress relaxation
time with a factor of $\sqrt{n}$, the relation between this quantity
and the CaBER relaxation time becomes linear (see figure
\ref{lambda_NWURZEL(n)_vs_Lambda_CaBER}).

The dependencies on the solvent viscosity are the
same for the CaBER and the normal stress relaxation time; this fact leads to the
result that the viscosity cancels out if one compares the first
normal stress coefficient with the CaBER relaxation time. Since the
scaling between $\lambda_C$ and $\lambda_N$ is simply the square
root of the number density per unit volume, the additional factor $n$
in equation \ref{Psi1} is canceled out too, meaning that we get a
quadratic dependency if we plot $\Psi_1$ versus $\lambda_C$ (figure
\ref{Psi1_vs_lambda_CaBER_gesamt}). It is worth mentioning that this
plot can be performed without any additional scaling, i.e., the
$\Psi_1$ data that are determined from the $N_1$ data are directly
plotted versus the $\lambda_C$ that are measured in the CaBER.

%--------------------------- SEC VI ------------------------
\section{Conclusion}
\label{sec6}
%-------------------------------------------------------

In conclusion, we have shown that the normal stress coefficient
$\Psi_1$ shows a quadratic dependence on the relaxation time that
we measured in the Capillary Breakup Extensional Rheometer. This can
be motivated by simple assumptions, i.e., by the fact the
polymer relaxation time that has been determined in in shear flow depends only weakly
on concentration while for the CaBER relaxation time an almost
linear dependence was found. This result could be obtained because
it was found that for long-chained, flexible polymers (PAAm, PEO),
one can quadratically fit the shear rate dependence of the first normal stress difference , as predicted by the Oldroyd-B model. From
this fit, we obtained the first normal stress coefficient $\Psi_1$.
The FENE-P showed no advantage to the simple Hookean elastic
dumbbell model (Oldroyd-B).

The dependence of the respective relaxation times on solvent
viscosity is roughly the same for CaBER and Rheometer data, i.e., it
is close to a linear dependence. Furthermore, the dependency on the
number of polymers for the two relaxation times differs by a factor
of $\sqrt{n}$. Scaling one relaxation time with this factor, one gets
a linear relationship to the other one. This $\sqrt{n}$ scaling
exactly explains the direct quadratic dependence of the first normal
stress coefficient on the CaBER relaxation time. The additional
linear factor $n$ in equation \ref{Psi1} is balanced by the
dependence of $\lambda_C$ on $n$. In summary, we showed that
there exists a very prominent and direct relationship between the
elastic fluid properties in a stationary shear experiment and the
parameters determined in a non-stationary extensional flow.


\begin{thebibliography}{10}
\expandafter\ifx\csname url\endcsname\relax
  \def\url#1{\texttt{#1}}\fi
\expandafter\ifx\csname urlprefix\endcsname\relax\def\urlprefix{URL }\fi
\expandafter\ifx\csname href\endcsname\relax
  \def\href#1#2{#2} \def\path#1{#1}\fi

\bibitem{Zimm56}
B.~Zimm, {Dynamics of polymer molecules in dilute solution: viscoelasticity,
  flow birefringence and dielectric loss}, J. Chem. Phys. 24
  (1956) 269.

\bibitem{Rouse53}
P.~Rouse~Jr., {A theory of the linear viscoelastic properties of dilute
  solutions of coiling polymers}, J. Chem. Phys. 21 (1953)
  1272.

\bibitem{Gupta00}
R.~Gupta, D.~Nguyen, T.~Sridhar, {Extensional viscosity of dilute polystyrene
  solutions: Effect of concentration and molecular weight}, Phys. Fluids
  12 (2000) 1296.

\bibitem{Sridhar91}
T.~Sridhar, V.~Tirtaatmadja, D.~Nguyen, R.~Gupta, {Measurement of extensional
  viscosity of polymer solutions}, J. Non-Newtonian Fluid Mech.
  40~(3) (1991) 271--280.

\bibitem{Tirtaatmadja93}
V.~Tirtaatmadja, T.~Sridhar, {A filament stretching device for measurement of
  extensional viscosity}, J. Rheol. 37 (1993) 1081.

\bibitem{Lindner03}
A.~Lindner, J.~Vermant, D.~Bonn, {How to obtain the elongational viscosity of
  dilute polymer solutions?}, Physica A: Statistical Mechanics and its
  Applications 319 (2003) 125--133.

\bibitem{Fuller87}
G.~Fuller, C.~Cathey, B.~Hubbard, B.~Zebrowski, {Extensional viscosity
  measurements for low-viscosity fluids}, J. Rheol. 31 (1987) 235.

\bibitem{Plog05}
J.~Plog, W.~Kulicke, C.~Clasen, {Influence of the molar mass distribution on
  the elongational behaviour of polymer solutions in capillary breakup},
  Appl. Rheol. 15~(1) (2005) 28--37.

\bibitem{Clasen06}
C.~Clasen, J.~Plog, W.~Kulicke, M.~Owens, C.~Macosko, L.~Scriven, M.~Verani,
  G.~McKinley, {How dilute are dilute solutions in extensional flows?}, J. Rheol. 50 (2006) 849.

\bibitem{Tirtaatmadja06}
V.~Tirtaatmadja, G.~McKinley, J.~Cooper-White, {Drop formation and breakup of
  low viscosity elastic fluids: Effects of molecular weight and concentration},
  Phys. Fluids 18 (2006) 043101.

\bibitem{Amarouchene01}
Y.~Amarouchene, D.~Bonn, J.~Meunier, H.~Kellay, {Inhibition of the finite-time
  singularity during droplet fission of a polymeric fluid}, Phys. Rev. Lett. 86~(16) (2001) 3558--3561.

\bibitem{Bird87}
R.~Bird, R.~Armstrong, O.~Hassager, {Dynamics of polymeric liquids, Vol. 1,
  Fluid mechanics}, John Wiley\& Sons, New York, 1987.

\bibitem{Schummer83}
P.~Sch{\"u}mmer, K.~Tebel, {A new elongational rheometer for polymer
  solutions}, J. Non-Newtonian Fluid Mech. 12~(3) (1983) 331--347.

\bibitem{Clasen06b}
C.~Clasen, J.~Eggers, M.~Fontelos, J.~Li, G.~McKinley, {The beads-on-string
  structure of viscoelastic threads}, J. Fluid Mech. 556 (2006)
  283--308.

\bibitem{Sattler08}
R.~Sattler, C.~Wagner, J.~Eggers, {Blistering pattern and formation of
  nanofibers in capillary thinning of polymer solutions}, Phys. Rev. Lett. 100~(16) (2008) 164502.

\bibitem{Renardy94}
M.~Renardy, {Some comments on the surface-tension driven break-up(or the lack
  of it) of viscoelastic jets}, J. Non-Newtonian Fluid Mech. 51~(1)
  (1994) 97--107.

\bibitem{Renardy95}
M.~Renardy, {A numerical study of the asymptotic evolution and breakup of
  Newtonian and viscoelastic jets}, J. Non-Newtonian Fluid Mech.
  59~(2-3) (1995) 267--282.

\bibitem{Rodd05}
L.~Rodd, T.~Scott, J.~Cooper-White, G.~McKinley, {Capillary break-up rheometry
  of low-viscosity elastic fluids}, Appl. Rheol. 15~(1) (2005) 12--27.

\bibitem{Oliveira05}
M.~Oliveira, G.~McKinley, {Iterated stretching and multiple beads-on-a-string
  phenomena in dilute solutions of highly extensible flexible polymers},
  Phys. Fluids 17 (2005) 071704.

\bibitem{Oliveira06}
M.~Oliveira, R.~Yeh, G.~McKinley, {Iterated stretching, extensional rheology
  and formation of beads-on-a-string structures in polymer solutions},
  J. Non-Newtonian Fluid Mech. 137~(1-3) (2006) 137--148.

\bibitem{Macosko94}
C.~Macosko, {Rheology: principles, measurements, and applications}, VCH:
  Wiley-VCH, New York, NY, 1994.

\bibitem{Rubinstein03}
M.~Rubinstein, R.~Colby, {Polymer physics}, Oxford University Press, USA, 2003.

\end{thebibliography}
\end{document}